\documentclass{aa}  

\usepackage{xcolor}
\usepackage{multicol}
\usepackage[titletoc]{appendix}
\usepackage{float}
\usepackage{caption}
\usepackage{subfig}
\usepackage{graphicx}
\usepackage{graphics}
\usepackage{titlesec}
\usepackage{cuted}
\usepackage{multirow,tabularx}
\usepackage{txfonts}
\renewcommand\arraystretch{1.4}
\renewcommand{\tabcolsep}{6pt}
\usepackage{natbib}
\bibpunct{(}{)}{;}{a}{}{,} 
\begin{document}

\title{Does the Fe L-shell blend bias abundance measurements in intermediate-temperature clusters?}


\author{Giacomo Riva
    \inst{1,2}
    \and
    Simona Ghizzardi
    \inst{1}
    \and
    Silvano Molendi
    \inst{1}
    \and
    Iacopo Bartalucci
    \inst{1}
    \and
    Sabrina De Grandi
    \inst{3}
    \and
    Fabio Gastaldello
    \inst{1}
    \and
    Claudio Grillo
    \inst{1,2}
    \and 
    Mariachiara Rossetti
    \inst{1}
    }

\institute{INAF - Istituto di Astrofisica Spaziale e Fisica Cosmica di Milano, via A. Corti 12, 20133 Milano, Italy\\
\email{giacomo.riva@inaf.it}
\and
Dipartimento di Fisica, Università degli Studi di Milano, via G. Celoria 16, 20133 Milano, Italy
\and
INAF – Osservatorio Astronomico di Brera, via E. Bianchi 46, 23807 Merate, Italy}


\date{Received XXX; accepted XXX}

 
\abstract
{In intermediate-mass ($M = 2-4 \times 10^{14}$ M$_{\odot}$, or equivalently $T = 2.5-4.5$ keV) galaxy clusters, abundance measurements are almost-equally driven by Fe K and L transitions, at $\sim 6.7$ keV and $0.9-1.3$ keV, respectively. While K-shell-derived measurements are considered reliable, the resolution of the currently available instrumentation, as well as our current knowledge of the atomic processes, makes the modelling of the L-line complex challenging, resulting in potential biases for abundance measurements. In this work, we study with unprecedented accuracy systematics related to the modelling of the Fe L-line complex that may influence iron-abundance measurements in the intermediate-mass range. 
To this aim, we select a sample of three bright and nearby galaxy clusters, with long XMM-\textit{Newton} observations available and temperature in the $2.5-4.5$ keV range. We fit spectra extracted from concentric rings with APEC and APEC+APEC models, by alternatively excluding the L and K bands, and derive the fractional difference of the metal abundances, $\Delta Z/Z$, as indication of the consistency between K- and L-shell-derived measurements. The $\Delta Z/Z$ distribution is then studied as a function of the cluster radius, ring temperature and X-ray flux. The L-blend-induced systematics, measured through an individual fit of each MOS and pn spectrum, remain constant at a $5-6\%$ value in the whole $2.5-4.5$ keV temperature range. Conversely, a joint fit of MOS and pn spectra leads to a slight excess of $1-2\%$ in the above estimate. No significant dependence on the ring X-ray flux is highlighted. The measured $5-8\%$ value indicates a modest contribution of the systematics to the derived iron abundances, giving confidence for future measurements. To date, these findings represent the best-achievable estimate of the systematics in analysis, while future microcalorimeters will significantly improve our understanding of the atomic processes underlying the Fe L emissions.}


\maketitle
%
\section{Introduction}
\begin{table*} 
\caption{Cluster properties and information of the XMM-\textit{Newton} observations.}
\centering                          
\begin{tabular}{cccccccc}        
\hline   
\hline 
Cluster & Obs. ID &MOS Exposure & pn Exposure  & $R_{500}$ & $T$ & Flux ($0.1-2.4$ keV) &$z$ \\
&&(ks)& (ks) &  (Mpc) & (keV)& (erg s$^{-1}$ cm$^{-2}$) & \\
\hline  
Perseus       & $0305780101$ & 118 & 94 &$1.3$&$3.5-4.5$ & $9.61  \times 10^{-10}$ & $0.018$ \\ 
2A 0335+096   & $0147800201$ & 79 & 76& $0.9$&$2.4-3.5$ & $9.16  \times 10^{-11}$ & $0.036$ \\
Sérsic 159-03 & $0147800101$ & 86 & 77 & $0.8$ & $2.5-2.7$ & $2.49 \times 10^{-11}$ & $0.056$ \\
\hline                                   
\end{tabular}
\tablefoot{"MOS Exposure" and "pn Exposure" are the effective exposure times of MOS and pn cameras, after the soft-proton cleaning. Estimates for $R_{500}$ are taken from \citet{giacintucci19} for Perseus and 2A 0335+096 and from \citet{sun09} for Sérsic 159-03. In the Table we list, for each cluster, the temperature ranges covered by the various regions for which we extracted and fitted the spectra. These temperatures lie in the range we are interested in. Regions with temperature higher than $4.5$ keV are not considered for Perseus, as explained in Sect. 3.2.1. Perseus flux and redshift are taken from \citeauthor{ebeling02} \citeyear{ebeling02}; for both 2A 0335+096 and Sérsic 159-03, we refer to \citeauthor{deplaa07} \citeyear{deplaa07} and \citeauthor{sanders11} \citeyear{sanders11} for the flux and the redshift, respectively. }             
\label{table:1}
\end{table*}

During the last decades, several studies have focussed on the measurement of the metallicity in galaxy clusters and groups (see \citeauthor{mernier18a} \citeyear{mernier18a} and \citeauthor{gastaldello21} \citeyear{gastaldello21} for a review). Since they are the largest structures to have clearly decoupled from the Hubble flow, studies of metal abundance in the Intra-Cluster Medium (ICM) provide important clues on the cluster formation and evolution processes, as well as on the enrichment of the Universe. Since the $1950$'s, when the first complete theories of stellar nucleosynthesis (\citeauthor{cameron57} \citeyear{cameron57}; \citeauthor{burbidge57} \citeyear{burbidge57}) were developed, it has been generally accepted that most of the chemical elements heavier than helium are produced in stars at different stages (particularly at the end) of their lifetimes (see \citeauthor{nomoto13} \citeyear{nomoto13} for a review). Although the synthesis processes of the heavy elements in the Universe are known fairly well, the general picture of the enrichment history in galaxy clusters and groups is still far from being completely clear. 

Galaxy-cluster outskirts have recently enjoyed particular attention, since they provide direct information on the ICM (its formation, accretion and contribution to the growth of large-scale structures; \citeauthor{walker19} \citeyear{walker19}). However, iron-abundance measurements are not easy to perform in clusters, especially in their outermost, low-surface-brightness regions. Reliable measurements have typically been carried out up to $R \sim  0.6\,R_{500}$\footnote{$R_{\Delta}$ is the radius inside which the cluster mass density is $\Delta$ times the critical density of the Universe $\rho_{\text{c}}(z)=3H_0^2 E(z)^2/8\pi G$, where $E(z)=\sqrt{\Omega_\text{m}(1+z)^3 + \Omega_{\Lambda}}$ (in a flat Universe) and $z$ is the redshift. $M_{\Delta}$ is then defined as the total mass within $R_{\Delta}$.}, sampling about one third of the total gas mass in clusters (as discussed in \citeauthor{molendi16} \citeyear{molendi16}) and only on archival samples without guarantee of representativity (\citeauthor{mernier17} \citeyear{mernier17}; \citeauthor{leccardi08a} \citeyear{leccardi08a}; \citeauthor{lovisari19} \citeyear{lovisari19}). In this context, a very-large programme on XMM-\textit{Newton} (the XMM-\textit{Newton} Cluster Outskirts Project, X-COP; \citeauthor{eckert17} \citeyear{eckert17}) has been carried out to advance our knowledge of the physical conditions in the cluster outskirts, pushing the analyses up to $R_{500}$, for a sample of 12 massive ($M_{500} >4\times 10^{14}$ M$_{\odot}$) galaxy clusters selected in the \textit{Planck} all-sky survey of Sunyaev-Zeldovich (SZ, \citeauthor{sunyaev72} \citeyear{sunyaev72}) sources (\citeauthor{planck14} \citeyear{planck14};  \citeauthor{planck15} \citeyear{planck15}). 

Within this programme, \citet{ghizzardi21} derived for the first time robust metallicity profiles up to $R_{500}$ for a sample of galaxy clusters and identified a main systematic error that affects iron-abundance measurements in the cluster outskirts. In the central regions of massive clusters (as for the X-COP sample, where temperatures are higher than $4$ keV), iron-abundance measurements are mostly driven by the Fe K transitions (mainly by Fe XXV and Fe XXVI K$\alpha$ lines) at $\sim 6.7$ keV in the ICM spectrum, while the Equivalent Width (EW) of the Fe L-line complex\footnote{Hereafter, we will refer to the just-mentioned transitions simply as K$\alpha$ and L-shell, respectively.} ($0.9-1.3$ keV) is smaller (see Fig. 3 of their work). However, going towards the cluster outskirts, the EW of the K$\alpha$ reduces with respect to the total continuum, which is dominated by the background emission, and the L-shell assumes a primary role in the fit (see Fig. 4 of their work). The L-line-derived abundance measurements in such high-background regime were found to be really puzzling in some cases, making the reliability of the L-shell in abundance measurements questionable. The exclusion of the L-shell ($0.9-1.3$ keV) energy band was found to lead to more-reliable abundance profiles, renewing some old but still relevant questions: are the L-shell-derived abundance measurements unreliable in general, or do they become so only in specific and complex situations? 

At the spectral resolution of currently available CCD cameras, the single lines that build up the L-shell (see \citeauthor{hwang97} \citeyear{hwang97}; \citeauthor{mazzotta98} \citeyear{mazzotta98}; \citeauthor{gu19} \citeyear{gu19}; \citeauthor{gu20} \citeyear{gu20}; \citeauthor{heuer21} \citeyear{heuer21}) are not resolved and are actually seen as a blend (see Fig. 1 in \citeauthor{gastaldello02} \citeyear{gastaldello02}). The large number of Fe emission lines peaking in the $0.9-1.3$ keV energy range, as well as our current knowledge of the atomic processes, makes the modelling of the L-shell challenging, resulting in potential biases. Conversely, the Fe K$\alpha$ at $\sim 6.7$ keV is relatively well understood, especially after the data release of the SXS spectrum of Perseus \citep{hitomi18}. In the late 1990's, \citet{hwang97} quantified possible L-shell-induced systematics, by comparing K$\alpha$- and L-blend-derived iron abundances for a few galaxy clusters (with ASCA observations) in the $2-4$ keV temperature range. Their research did not highlight any significant discrepancies between the K- and L-derived measurements, within the $90\%$ confidence level. However, the large error bars of their results do not allow a precise assessment of the reliability of the L-shell-derived abundance measurements. 

Currently, we can rely on satellites with advanced characteristics (e.g. the large effective area of XMM-\textit{Newton}) and also on atomic codes (e.g. AtomDB\footnote{http://www.atomdb.org.}, implemented in the APEC model as part of the XSPEC\footnote{https://heasarc.gsfc.nasa.gov/xanadu/xspec.} fitting package; \citeauthor{smith01} \citeyear{smith01}; \citeauthor{foster12} \citeyear{foster12}; SPEXACT, available in the SPEX\footnote{https://www.sron.nl/astrophysics-spex.} fitting package; \citeauthor{kaastra96} \citeyear{kaastra96}; \citeauthor{kaastra18} \citeyear{kaastra18}) which describe the atomic processes that produce the emission lines observed in the ICM spectra with greater precision than in the past. These instruments allow us to study with high accuracy systematics related to the L-shell that affect iron-abundance measurements. In this work, we compare K$\alpha$- and L-shell-derived abundance measurements for a sample of galaxy clusters observed with XMM-\textit{Newton} and with temperatures in the $2.5-4.5$ keV range. In this range, the EWs of the K$\alpha$ and L-shell (at these temperatures primarily due to Fe XXIII and XXIV) emissions are comparable, giving us the possibility to test the atomic codes underlying abundance measurements. 

The paper is organised as follows: in Sect. 2, we present our sample and the data reduction; Section 3 is dedicated to the X-ray data analysis: here we define the strategies that are used to estimate the L-shell-induced systematics; in Sect. 4 we analyse a possible dependence of the systematics on the temperature and the flux, while Section 5 is dedicated to the discussion of the results; finally, in Sect. 6 we summarise our main findings.

Throughout the paper, we assume a $\Lambda$ cold-dark-matter cosmology with $H_0 = 70$ km s$^{-1}$ Mpc$^{-1}$, $\Omega_{\text{m}} = 0.3$, $\Omega_{\Lambda} = 0.7$ and $E(z) =\sqrt{\Omega_{\text{m}}(1 + z)^3 + \Omega_{\Lambda}}$ for the evolution of the Hubble parameter. The solar abundance table is set to \citet{asplund09}, while all the quoted errors hereafter are at the $1\sigma$ confidence level.

\section{Selection of the sample and data processing}

In this work, we restrict our analysis to a sample of nearby and bright clusters, with i) long XMM-\textit{Newton} observations ($\gtrsim 120$ ks), ii) good statistics, iii) regions with temperature in the $2.5-4.5$ keV range and iv) high source-to-background (S/B) ratio, to focus on the contribution to the systematics related to the underlying atomic model in the above temperature range and to neglect contributions from the background contamination. To this aim, we select the following galaxy clusters: Perseus, 2A 0335+096 and Sérsic 159-03 (Table \ref{table:1}).

For each cluster, we only consider the longest available observation in the XMM-\textit{Newton} Science Archive\footnote{https://www.cosmos.esa.int/web/xmm-newton/xsa.} (XSA), to avoid additional systematics deriving from a combination of different observations of the same cluster. The identification numbers of the selected observations are listed in Table \ref{table:1}, together with the cluster redshifts, temperatures and X-ray fluxes in the $0.1-2.4$ keV band. Given the redshifts of the three clusters (see Table \ref{table:1}), $1'$ corresponds to 22 kpc, 43 kpc and 65 kpc for Perseus, 2A 0335+096 and Sérsic 159-03, respectively.

X-ray data are reduced using the XMM-\textit{Newton} Science Analysis Software\footnote{https://www.cosmos.esa.int/web/xmm-newton/sas.} (SAS). The main steps of the data processing are the following: 
\begin{itemize}
    \item we start by producing calibrated event files for each observation (with \textit{emchain} and \textit{epchain} tasks);
    \item we extract light curves in the hard ($10-12$ keV for MOS detectors and $10-13$ keV for pn) band by using a $100$-s binning. We filter out time intervals affected by soft-proton flares by applying the appropriate threshold for each instrument. For MOS detectors we adopt a threshold of $0.25$, $0.17$ and $0.16$ cts s$^{-1}$ for Perseus, 2A 0335+096 and Sérsic 159-03, respectively, while the threshold is set to $0.70$, $0.43$, $0.42$ cts s$^{-1}$ for the three pn light curves;
    \item following \citet{leccardi08b}, we then reject time periods in which count rates exceed $3\sigma$ from the median of the light curve in the $2-5$ keV band, in order to eliminate contributions of softer flares. The effective exposure times of MOS and pn cameras after the soft-proton cleaning are listed in Table \ref{table:1};
    \item finally, we select by eye and excise the bright point sources inside the regions from which spectra are extracted (see next section).
\end{itemize}

A characterisation of the local sky background cannot be considered for Perseus, 2A 0335+096 and Sérsic 159-03, since their thermal emissions fill the whole XMM-\textit{Newton} FoV. For this reason and since we are interested in studying the L-shell systematics in condition of relatively high surface brightness, where the background subtraction is less critical than for external low-surface-brightness regions, we use blank-sky fields instead of proceeding with a more-detailed modelling of the different background components. In Sect. 4.1 we investigate the impact of this choice on our results. The blank-sky fields for EPIC MOS and pn were produced by \citet{leccardi08b}. We perform a background rescaling for each observation to account for temporal variations of the background (\citeauthor{leccardi08b} \citeyear{leccardi08b}; \citeauthor{ghizzardi14} \citeyear{ghizzardi14}).

\section{Data analysis}

\subsection{Spectral fitting}
\begin{figure}
    \centering
    \subfloat{\includegraphics[height=0.2\textwidth]{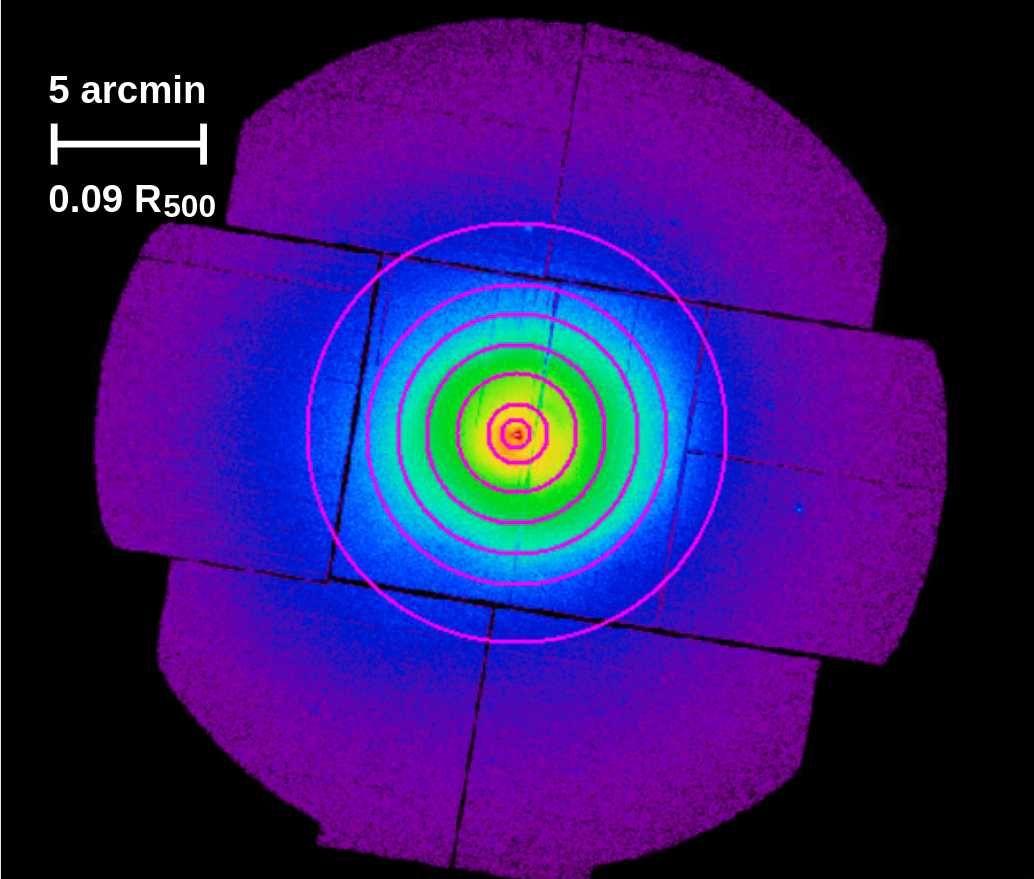}} \hfill
    \subfloat{\includegraphics[height=0.2\textwidth]{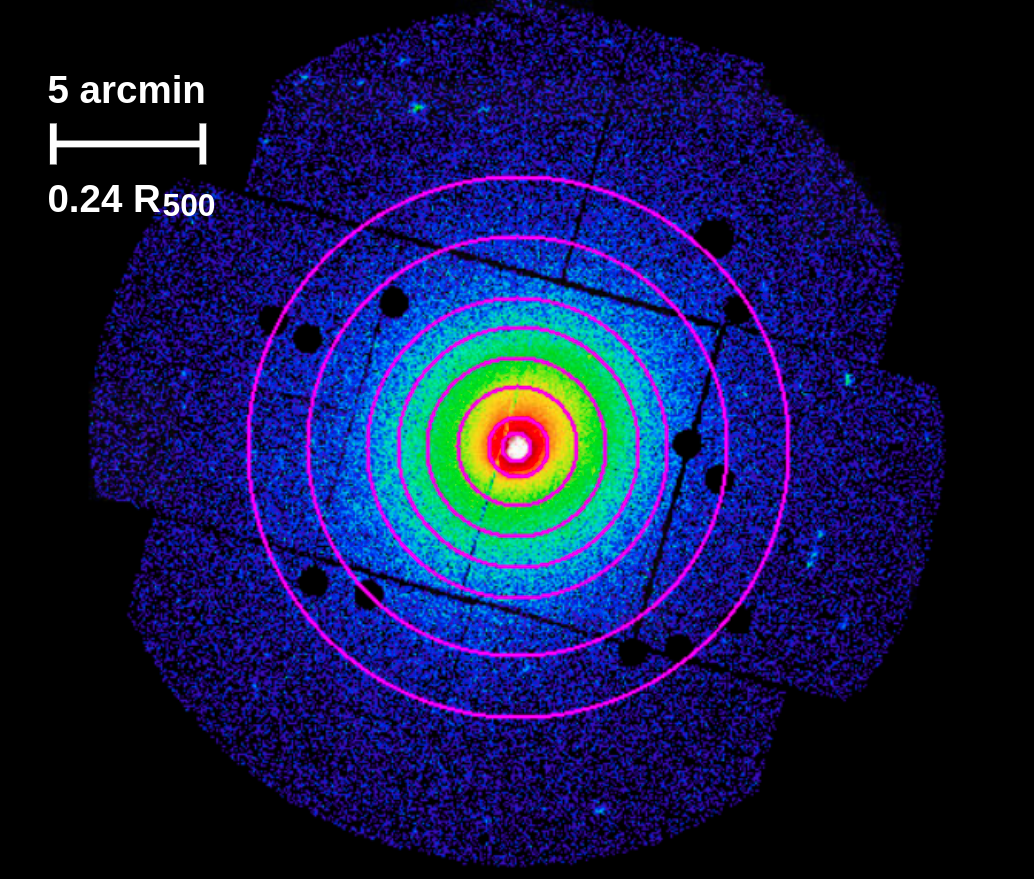}} \hfill
    \subfloat{\includegraphics[height=0.2\textwidth]{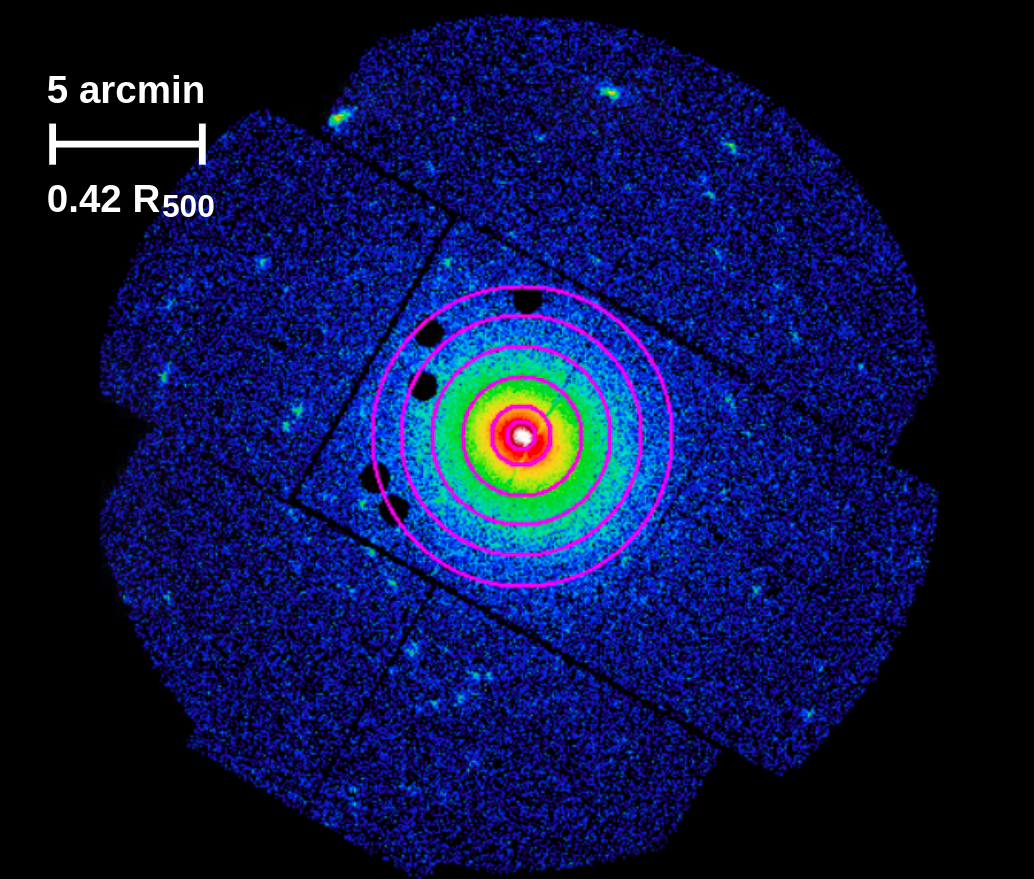}} 
    \caption{Surface-brightness map (MOS 2 camera) of Perseus (top left), 2A 0335+096 (top right) and Sérsic 159-03 (bottom), extracted in the $0.15-12$ keV band. X-ray point sources selected during the data reduction are excised. Rings from which the X-ray spectra are extracted are shown in magenta. Given the redshifts of the three clusters (see Table \ref{table:1}), $1'$ corresponds to 22 kpc, 43 kpc and 65 kpc for Perseus, 2A 0335+096 and Sérsic 159-03, respectively.}
    \label{ds9}
\end{figure}

We extract spectra from concentric rings (with centre located at the X-ray emission peak), to derive radial abundance profiles for each cluster. Spectra are extracted up to $7'$, $9'$ and $5'$ for Perseus, 2A 0335+096 and Sérsic 159-03, respectively (Fig. \ref{ds9}), according to the following pattern: rings are i) $0.5'$-wide within $1'$, ii) $1'$-wide up to $5'$, and iii) $2'$-wide at larger radii. Given the statistics of the three selected clusters, the above ring angular dimensions allow to derive temperature and abundance radial profiles with good precision. As detailed below, the external limit for Perseus is set by the high temperature at large radii, while in the case of 2A 0335+096 and Sérsic 159-03 we are limited by the presence of an increasingly dominating background. 

Spectra are fitted using XSPEC v.12.9.1p \citep{arnaud96}. At first, MOS and pn spectra of each region are fitted separately. We then perform a joint fit of the three cameras. The spectra are grouped to ensure a minimum of 25 counts per spectral channel. Unless otherwise stated, the energy range considered in the fitting is $0.7-10$ keV. We adopted the following models:
\begin{itemize}
    \item an absorbed (phabs) single-temperature APEC model (hereafter called "1T model"), with temperature, Galactic hydrogen column density $N_{\text{H}}$, abundance, redshift and normalisation free to vary;
    \item an absorbed (phabs) double-temperature APEC+APEC model, with two abundances linked together (hereafter "2T 1Z model"). Also in this case, temperatures, Galactic hydrogen column density, abundance, redshift and normalisations are free to vary. 
\end{itemize}
As further addressed in Sect. 3.2.2, the single-temperature ("1T") model is adequate for cluster rings that can be considered nearly isothermal, while the double-temperature ("2T") model better describes the data in case of multiphase plasma. The 2T model is also useful to mimic and describe cases where multiphaseness is artificial, due to projection or to contamination from adjacent rings. For this reason we do not correct data for PSF nor deproject profiles and use the 2T model to take into account these effects.

\subsection{Estimate of the systematic error}

The estimate of the systematic error linked to the Fe L-shell follows three major steps, that will be addressed in the following subsections:
\begin{enumerate}
    \item we exclude cluster rings whose properties are not adequate for our purpose;
    \item for each spectrum, we then select the model that better describes the data;
    \item finally we define the strategies for the estimate of the systematics.
\end{enumerate}

\subsubsection{Selection of the rings}
As already discussed, the aim of this work is the study of systematics related to the L-shell at intermediate temperatures ($T=2.5-4.5$ keV) and in condition of negligible background contamination. For this reason, starting from the general ring selection defined at the beginning of Sect. 3.1, we take the following measures:
\begin{itemize}
     \item the innermost central bin ($0'-0'.5$) is not considered for any cluster. Projection effects play a significant role in the central regions, particularly so in the innermost bin, where the temperature gradient becomes very steep: here neither the 1T nor the 2T 1Z models adequately describe the spectra in this circumstance. Since our focus is the study of L-shell-induced systematics rather than modelling the temperature distribution in such a central region, for each cluster we reject the central bin from the analysis. Moreover, Perseus and Sérsic 159-03 are experiencing in their cores a phase of AGN feedback (\citeauthor{gendron-marsolais16} \citeyear{gendron-marsolais16}; \citeauthor{werner11} \citeyear{werner11}; \citeauthor{farage12} \citeyear{farage12}), which dominates the X-ray emission in the innermost region. 
     \item radial bins whose temperature exceeds the range of interest are excluded from the analysis. This requirement places limits only on Perseus, whose temperature exceeds $5$ keV beyond $3'$;
     \item radial bins where the background emission is comparable to the source emission in the K$\alpha$ region ($\sim 6.7$ keV) are excluded from the analysis. This choice places limits on 2A 0335+096 and Sérsic 159-03, where radii larger than $5'$ and $3'$, respectively, are not considered.
\end{itemize}

After the above considerations, we focus on the spectra extracted from the following rings:
\begin{itemize}
    \item $0'.5-1'$, $1'-2'$, $2'-3'$ for Perseus;
    \item $0'.5-1'$, $1'-2'$, $2'-3'$, $3'-4'$, $4'-5'$ for 2A 0335+096;
    \item $0'.5-1'$, $1'-2'$, $2'-3'$ for Sérsic 159-031.
\end{itemize}

\subsubsection{Selection of the model}
In order to select which model provides the best description of the data for each spectrum, we fit all the spectra with both the 1T and the 2T 1Z models. As a first step, we qualitatively analyse how the residuals (i.e. the ratio between the data and the model) vary, using the two models, at high energy (beyond 5 keV) and around the L-shell and the K$\alpha$ energy bands. The presence of an excess in the residuals at high energy, as well as a tension between the residuals around the L-shell and the K$\alpha$, is a clear indication that a double-temperature model is needed in the fitting. This qualitative inspection allows us to assess that the 2T 1Z model is required at small radii ($r<3'$, $<2'$ and $<2'$ for Perseus, 2A 0335+096 and Sérsic 159-03, respectively), while the 1T model correctly describes 2A 0335+096 and Sérsic 159-03 data at external radii. 

In order to quantitatively support our preliminary inferences, we compare the $\chi^2/$d.o.f. values obtained for each fit. Such comparison confirms that, for inner radii, the double-temperature model is needed, while for external radii the clusters in our sample can be considered nearly isothermal.

\subsubsection{Estimate of the systematics}

To estimate the systematic error related to the L-shell, we follow two slightly different strategies. We fit each spectrum with the selected (1T or 2T 1Z, see previous section) model, excluding the L-shell ($0.9-1.3$ keV) and the K$\alpha$ ($6.3-6.9$ keV, $6.2-6.8$ keV and $6.1-6.7$ keV\footnote{The K$\alpha$ energy band slightly changes with the cluster redshift.} for Perseus, 2A 0335+096 and Sérsic 159-03, respectively) bands alternatively, with temperature(s):
\begin{enumerate}
    \item free to vary (hereafter, this will be referred to as the "$1^{\text{st}}$ strategy");
    \item frozen to the best-fit values derived by fitting over the whole $0.7-10$ keV energy range (hereafter, this will be referred to as the "$2^{\text{nd}}$ strategy").
\end{enumerate}

In this way, the iron abundance can be measured through the L-shell and the K$\alpha$ emissions separately, providing an estimate of the difference between the two. Since the physics underlying the K$\alpha$ emission is simpler and better understood than that for the L-shell blend (as discussed in the introduction), we shall attribute the above difference to a systematic error in the L-shell measurement. This error is quantified as:
\begin{equation}
    \frac{\Delta Z}{Z} = \frac{Z_{\text{L-shell}}-Z_{\text{K$\alpha$}}}{Z_{\text{K$\alpha$}}},
\end{equation}
where $Z_{\text{L-shell}}$ and $Z_{\text{K}\alpha}$ are the iron abundances measured after excluding the K$\alpha$ and the L-shell energy bands, respectively. Here, $Z_{\text{K}\alpha}$ is considered as a reference in the iron-abundance measurement. The values obtained for $Z_{\text{K}\alpha}$ and $Z_{\text{L-shell}}$ for each cluster and for both the strategies are listed in Appendix A.
\begin{figure}
   \subfloat{\includegraphics[height=0.3\textwidth]{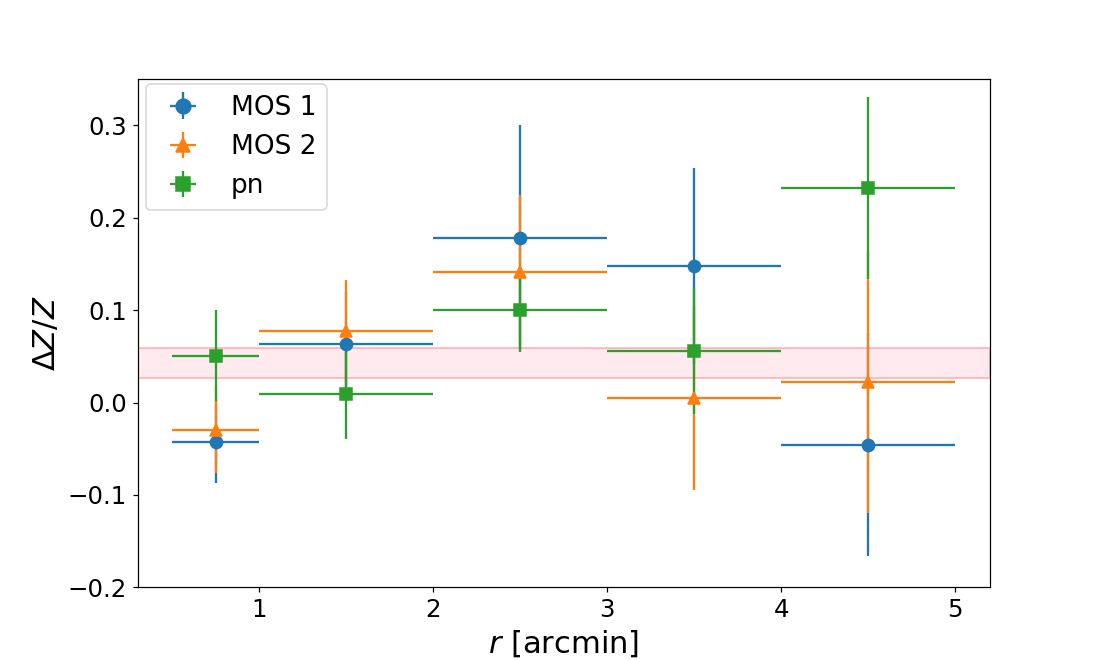}} \hfill
   \subfloat{\includegraphics[height=0.3\textwidth]{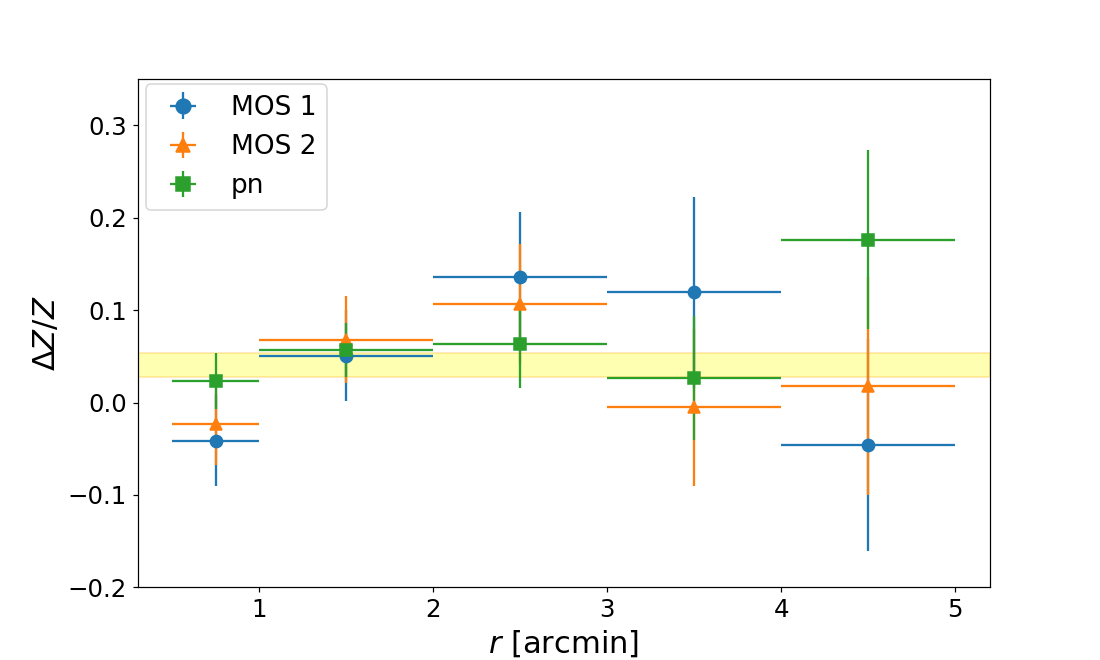}}\hfill
   \subfloat{\includegraphics[height=0.3\textwidth]{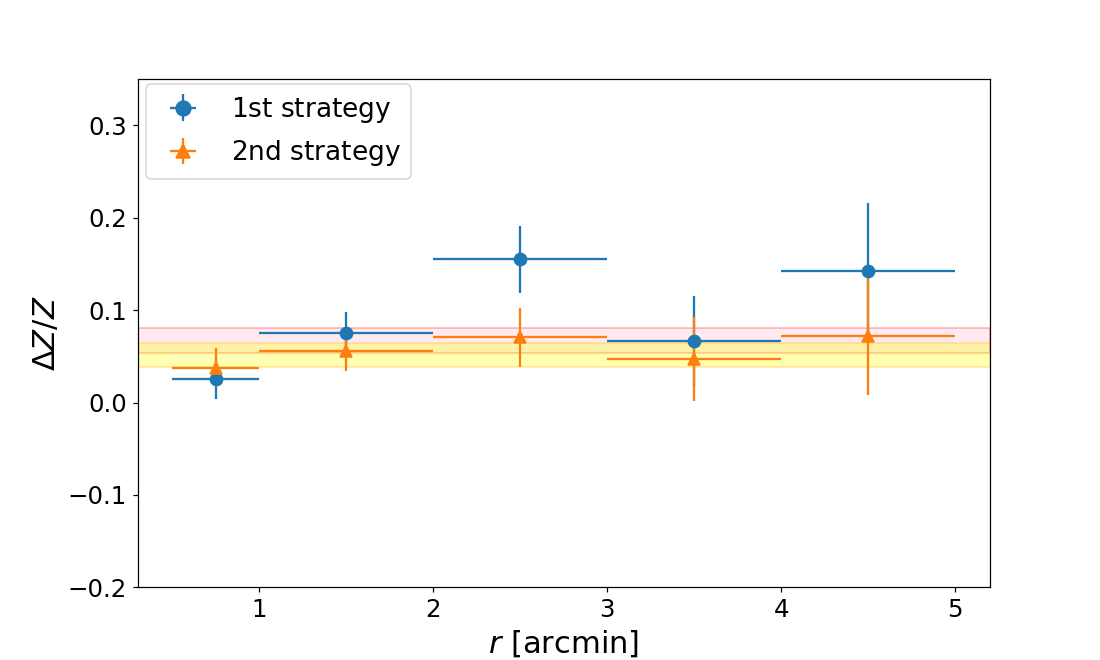}}
   \caption{Radial profile of the systematic error $\Delta Z/Z$ for 2A 0335+096. In the top and the central panels, we show the results obtained through an individual fit of the MOS and pn spectra of each region, following the $1^{\text{st}}$ and the $2^ {\text{nd}}$ strategies, respectively. In the bottom panel, we show the same profiles, with the MOS and pn spectra of each region considered jointly in the fitting. The weighted averages of the results are shown, together with their $1\sigma$ errors, as the pink and the yellow regions, for the $1^{\text{st}}$ and the $2^ {\text{nd}}$ strategies, respectively.}
   \label{2A}
\end{figure}

The above strategies are applied to each cluster of the sample and systematics are studied as a function of the cluster radii. As an example, in Fig. \ref{2A} (top and centre) we show the results obtained for 2A 0335+096, following the $1^{\text{st}}$ and the $2^ {\text{nd}}$ strategies and by separately fitting MOS and pn spectra; values obtained through a joint fit of the three detector spectra are displayed in Fig. \ref{2A} (bottom). 
Figure \ref{2A} shows that the systematic error $\Delta Z/Z$ related to the L-shell does not significantly vary with the cluster radius and can be considered constant, both looking at the individual and the joint fits of the MOS and pn spectra. In bright galaxy clusters, where statistical errors become tiny, the abundances measured by individually fitting MOS and pn data are usually found to be  significantly different, due to the well-known cross-calibration issues of the three XMM-\textit{Newton} detectors \citep{read14}. However, and interestingly, MOS and pn cameras lead to $\Delta Z/Z$ values consistent with each other, within the statistical errors, as shown in Fig. \ref{2A}. The results obtained for Perseus and Sérsic 159-03 are quite similar and displayed in Appendix B.

The weighted averages of the results obtained with both the individual and the joint fits for each cluster are shown in Table \ref{table:2}. The average values derived considering the three detectors together are indicated as "EPIC", while those derived separately are indicated as "MOS 1", "MOS 2" and "pn". The total ("EPIC") averages for 2A 0335+096, obtained through the $1^{\text{st}}$ and the $2^{\text{nd}}$ strategies with both the individual and the joint fits, are also shown, with the corresponding $1\sigma$ errors, as the pink and the yellow regions (respectively) in Fig. \ref{2A}.  
\begin{table*}
\caption{Weighted-averaged $\Delta Z/Z$ values, estimated for Perseus, 2A 0335+096 and Sérsic 159-03.} 
\centering
\begin{tabular}{ c|c|cc|cc  }
\hline
\hline
Cluster&Detector&\multicolumn{2}{c}{Individual fit}& \multicolumn{2}{c}{Joint fit} \\
\cline{3-6}
&&$1^{\text{st}}$ strategy & $2^{\text{nd}}$ strategy&$1^{\text{st}}$ strategy & $2^{\text{nd}}$ strategy\\
\hline
&MOS 1&$0.056\pm0.011$&$0.027\pm0.011$&$-$&$-$\\
Perseus& MOS 2&$0.044\pm 0.016$&$0.048\pm 0.012$&$-$&$-$\\
&pn&$0.076\pm 0.009$&$0.073\pm 0.010$&$-$&$-$\\
&EPIC&$0.064\pm0.007$ & $0.051 \pm 0.006$ & $0.081 \pm 0.006$ & $0.062 \pm 0.006$\\
\hline
&MOS 1&$0.020\pm 0.031$&$0.033\pm 0.029$&$-$&$-$\\
2A 0335+096& MOS 2&$0.030\pm 0.030$&$0.032\pm 0.027$&$-$&$-$\\
&pn&$0.067\pm 0.025$&$0.048\pm 0.018$&$-$&$-$\\
&EPIC&$0.043\pm0.016$ & $0.041 \pm 0.013$ & $0.068 \pm 0.014$ & $0.052 \pm 0.013$\\
\hline
&MOS 1&$-0.033\pm 0.070$&$-0.031\pm 0.063$&$-$&$-$\\
Sérsic 159-03& MOS 2&$0.112\pm 0.066$&$0.128\pm 0.065$&$-$&$-$\\
&pn&$-0.020\pm 0.046$&$0.039\pm 0.043$&$-$&$-$\\
&EPIC&$0.010\pm0.033$ & $0.042 \pm 0.031$ & $0.042 \pm 0.032$ & $0.048 \pm 0.029$\\
\hline
\end{tabular}
\tablefoot{$\Delta Z/Z$ values are obtained both considering the three XMM-Newton camera independently (indicated with "MOS 1", "MOS 2" and "pn") and together ("EPIC") in the average (see text).}
\label{table:2}
\end{table*}
Looking at the values listed in Table \ref{table:2}, we notice that the weighted averages of $\Delta Z/Z$ suggest a slightly increasing trend from Sérsic 159-03 to Perseus. However, large error bars do not allow to make further considerations and the systematic error estimated for each cluster is consistent with being constant at the same ($\sim 5-6\%$ and $\sim 6-8\%$ for the individual and the joint fits, respectively) value. 
The joint fits are found to lead to a slight excess of $\sim 1-2\%$ in the estimate of the systematics linked to the L-shell, with respect to the results obtained with the detectors considered separately. This excess is likely to be related to the above mentioned cross-calibration issues of the three XMM-Newton detectors \citep{read14}, resulting in an additional systematic error.


\section{Dependence on temperature and flux}

\subsection{Systematics vs Temperature}

\begin{figure*}
            \subfloat{\includegraphics[height=0.3\textwidth]{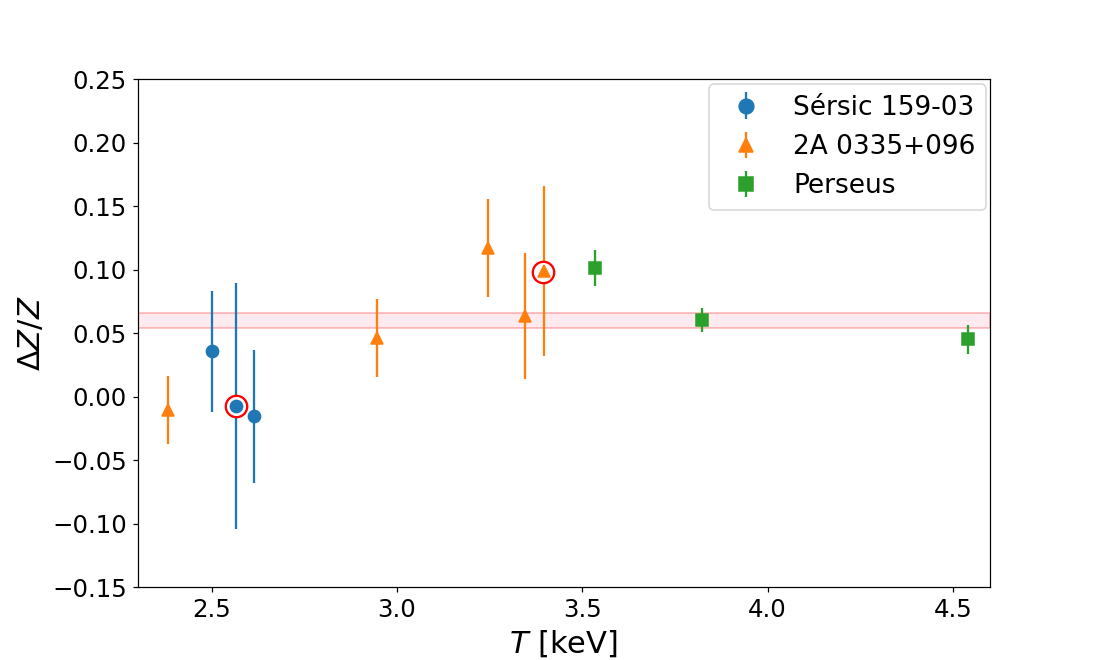}} \hfill
            \subfloat{\includegraphics[height=0.3\textwidth]{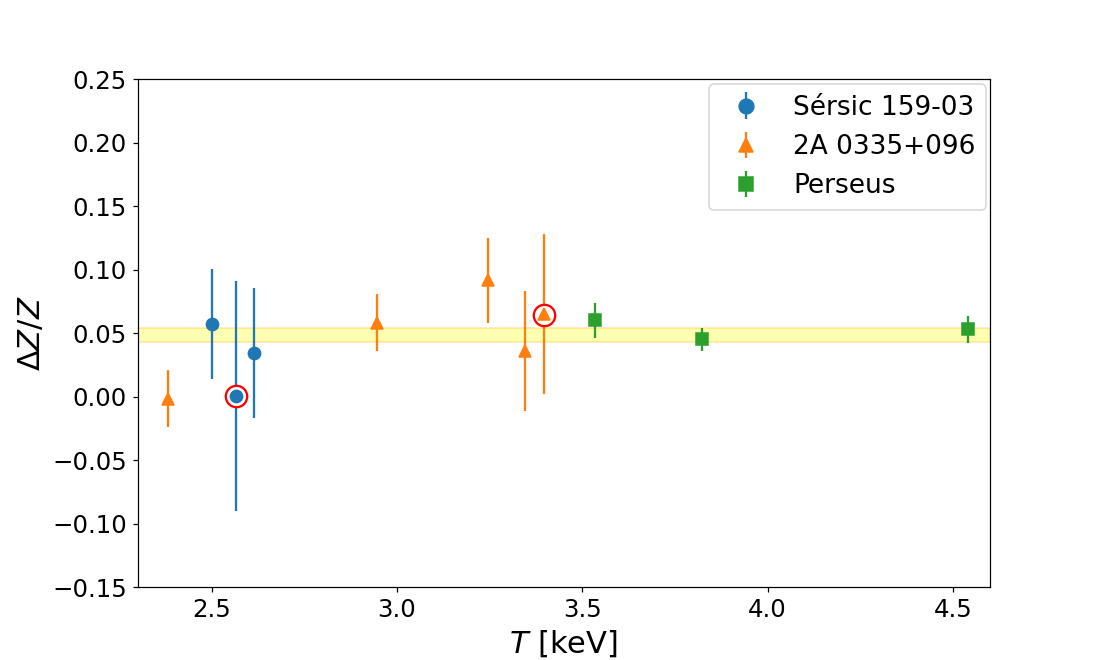}}
            \hfill
            \subfloat{\includegraphics[height=0.3\textwidth]{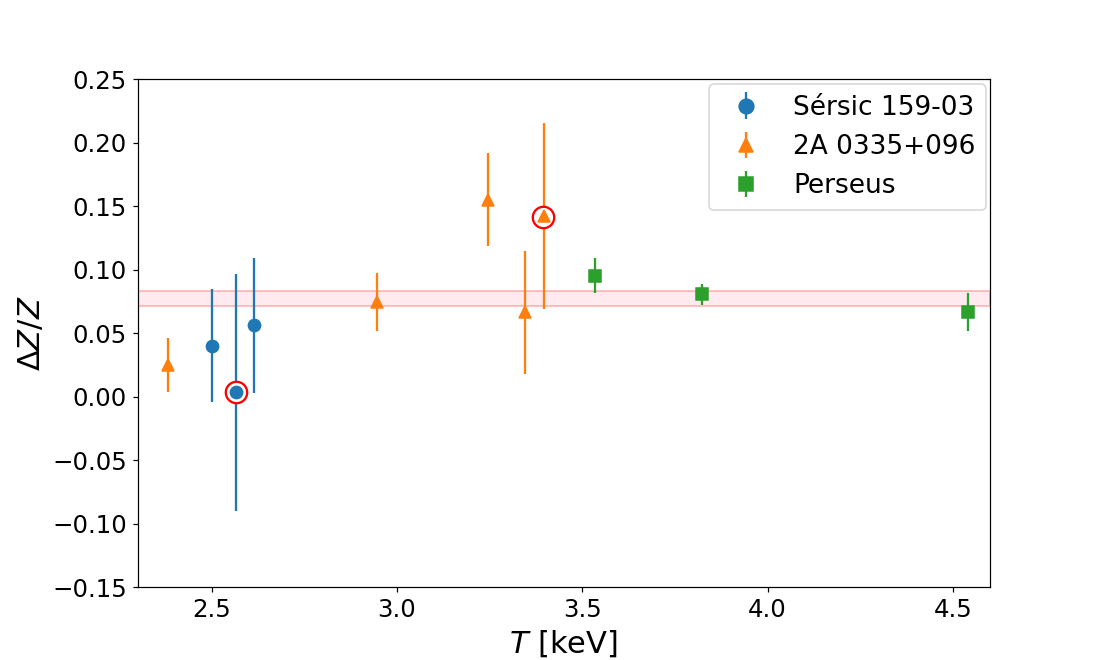}} \hfill
            \subfloat{\includegraphics[height=0.3\textwidth]{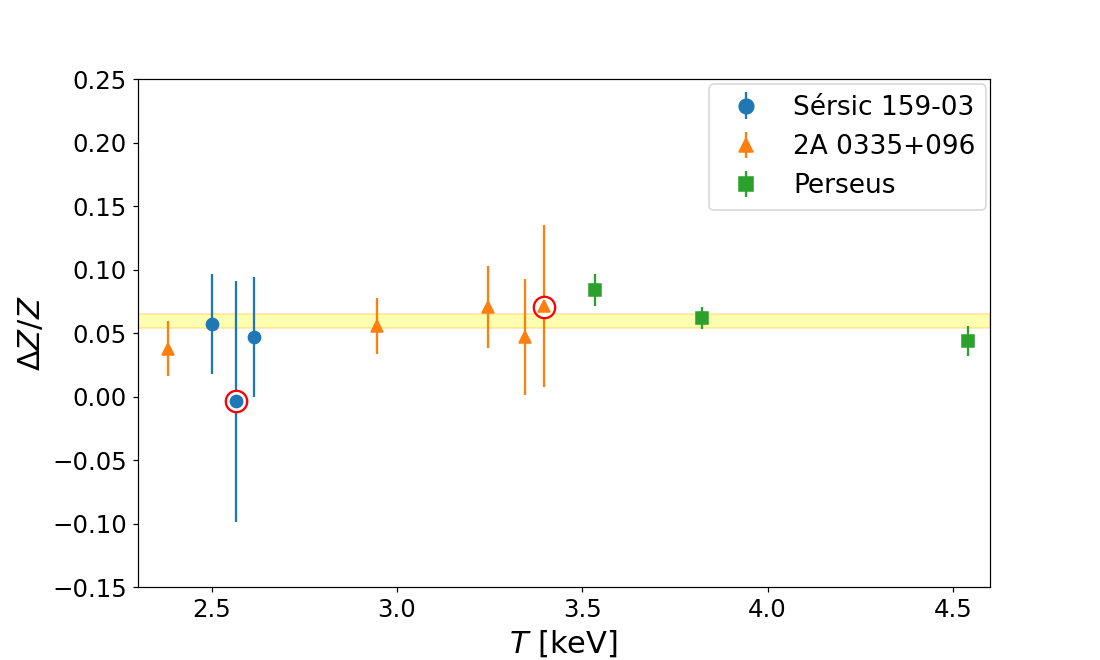}}
      \caption{Systematic error $\Delta Z/Z$ linked to the L-shell as a function of the temperature. The top and the bottom rows refer to the different fitting option (individual and joint fit of the MOS and pn spectra of each region, respectively), while the left and the right columns refer to the $1^{\text{st}}$ and the $2^{\text{nd}}$ strategies, respectively. The pink and the yellow regions represent the weighted averages of the results, together with their $1\sigma$ errors.}
         \label{Temp}
\end{figure*}

The analysis, independently run on the three clusters of our sample, has shown that the systematics related to the L-shell remain constant, both considering different radii of the same cluster (see Fig. \ref{2A}) and the three clusters themselves (see Table \ref{table:2}).
Since each cluster ring has its own temperature, the above result also gives us an indication that systematics seem not to vary with the temperature in the considered range and to be constant with a $\Delta Z/Z \sim5-6\%$ (or at least not higher than $8\%$) value. In order to study the dependence, if any, of the systematic error $\Delta Z/Z$ on the temperature, we use the weighted average of $\Delta Z/Z$ (averaged over the three detectors) for each cluster ring. The obtained values are plotted in Fig. \ref{Temp} (top left and right, for the $1^{\text{st}}$ and the $2^{\text{nd}}$ strategies respectively), as a function of the ring temperatures. The temperatures considered in this part of the analysis are obtained for each ring through the 1T model\footnote{An alternative way of calculating the temperatures for the rings needing the 2T 1Z model is to consider a weighted-averaged temperature on the normalisations of the APEC+APEC model. Such alternative way does not significantly modify the measured temperatures.}, with independent fits for each XMM-Newton camera spectrum, and then averaged over the three detectors\footnote{The gap between the temperatures measured for the $4^{\text{th}}$ and $5^{\text{th}}$ bins of 2A 0335+096 has been slightly widened, to avoid an overlap of the results.}. 

The three clusters of our sample allow to homogeneously study the $2.5-4.5$ keV range. Perseus gives information about the highest temperatures, i.e. the $3.5-4.5$ keV range. 2A 0335+096 allows to study the intermediate $3.0-3.5$ keV range, while Sérsic 159-03 (together with the innermost ring for 2A 0335+096) gives an estimate of the systematic error $\Delta Z/Z$ at lower temperatures. An inspection of Fig. \ref{Temp} (top panels) shows that the systematic error $\Delta Z/Z$ linked to the L-shell remains constant over the studied temperature range. The average values of the results (also reported, together with their $1\sigma$ error bars, as the pink and the yellow regions in Fig. \ref{Temp}, top panels), for the $1^{\text{st}}$ and the $2^{\text{nd}}$ strategies, respectively, are the following: 
\begin{enumerate}
    \item $\Delta Z/Z = 0.060 \pm 0.006$;
    \item $\Delta Z/Z = 0.049 \pm 0.005$.
\end{enumerate}
Considering MOS and pn spectra separately in the fitting leads to a $\sim 5-6\%$ systematic error associated to the Fe L-shell, in the whole $2.5-4.5$ keV temperature range. The result obtained through the $2^{\text{nd}}$ strategy is slightly lower than that derived through the $1^{\text{st}}$ one and they are consistent with each other within their error bars. This finding suggests that temperatures free to vary do not significantly affect our results and seem not to introduce additional systematics in the iron-abundance measurements. 

We then repeat the same estimate, this time considering a joint fit of the three camera spectra. The results are shown in Fig. \ref{Temp} (bottom left and right, for the $1^{\text{st}}$ and the $2^{\text{nd}}$ strategies, respectively). Also the $\Delta Z/Z$ values estimated through the joint fit remain constant over the studied temperature range, with values:
\begin{enumerate}
    \item $\Delta Z/Z = 0.078 \pm 0.006$;
    \item $\Delta Z/Z = 0.060 \pm 0.005$,
\end{enumerate}
for the $1^{\text{st}}$ and the $2^{\text{nd}}$ strategies, respectively. As already pointed out, a joint fit of the MOS and pn spectra leads to a $\sim1-2\%$ excess in the estimate of the systematic error linked to the L-shell, with respect to an individual fit of each spectrum. We already suggested that this behaviour is likely to be related to a cross-calibration issue of the three detectors \citep{read14}, leading to an overestimate of the systematics we are interested in. 

Although our study is mainly focussed on bright cluster regions, where the background contribution to the total emission is negligible, two of the considered rings (one for 2A 0335+096 and one for Sérsic 159-03, marked with red circles in Fig. \ref{Temp}) also allow us to explore regions where the background emission starts to play a role in the K$\alpha$ region (S/B $\sim 5$ and $\sim 3$ for the most external ring of 2A 0335+096 and Sérsic 159-03, respectively, in the $6-7$ keV range). Interestingly, as we see in Fig. \ref{Temp}, the systematics measured for such rings are in agreement, within their $1\sigma$ errors, with the average values. This finding suggests that a moderate contamination by the background emission seems not to lead to additional contributions to the systematics we are interested in. 

At this point, one may wonder if using blank sky fields instead of a detailed description of each background component leads to significantly biased results, especially for the most external rings of 2A 0335+096 and Sérsic 159-03, where the S/B ratio is smaller. To investigate further this issue, we refit spectra assuming a possible over- and under-estimation of the background of $10\%$, in order to evaluate potential systematics introduced by an incorrect estimate of the background contribution\footnote{A $10\%$ variation of the background contribution is above the current uncertainties related to the blank sky fields, which are considered to be about $5\%$. However, this is useful to place strong limits on background-related systematics.}. The $\Delta Z/Z$ values measured after modifying the background emission (and through the $2^{\text{nd}}$ strategy; top right panel in Fig. \ref{Temp}) show little scatter ($\pm 1\%$ and $\pm 3\%$ for the most external rings of Sérsic 159-03 and 2A 0335+096, respectively, well below the $\pm 9\%$ and $\pm 7\%$ statistical uncertainties), with respect to the standard (non-modified) value. It should be noticed that these tiny variations represent an upper limit to background-related effects on our results, since central rings are less sensitive to moderate variations of the background contamination, given their significantly higher S/B ratio. The above considerations show that the way we treated the background emission does not significantly influence our results and is sufficient for our purpose. However, in order to reliably estimate the systematics linked to the L-shell in conditions with stronger background emission, a precise description of the different background components should be considered.

\subsection{Systematics vs Flux}
\begin{figure*}
            \subfloat{\includegraphics[height=0.3\textwidth]{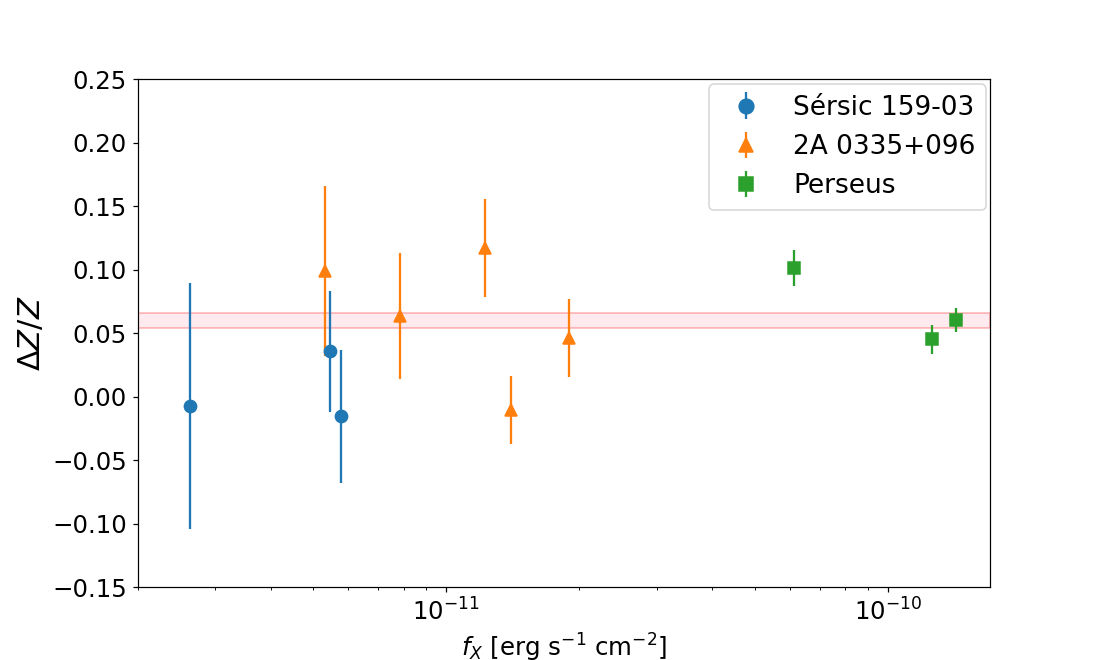}} \hfill
            \subfloat{\includegraphics[height=0.3\textwidth]{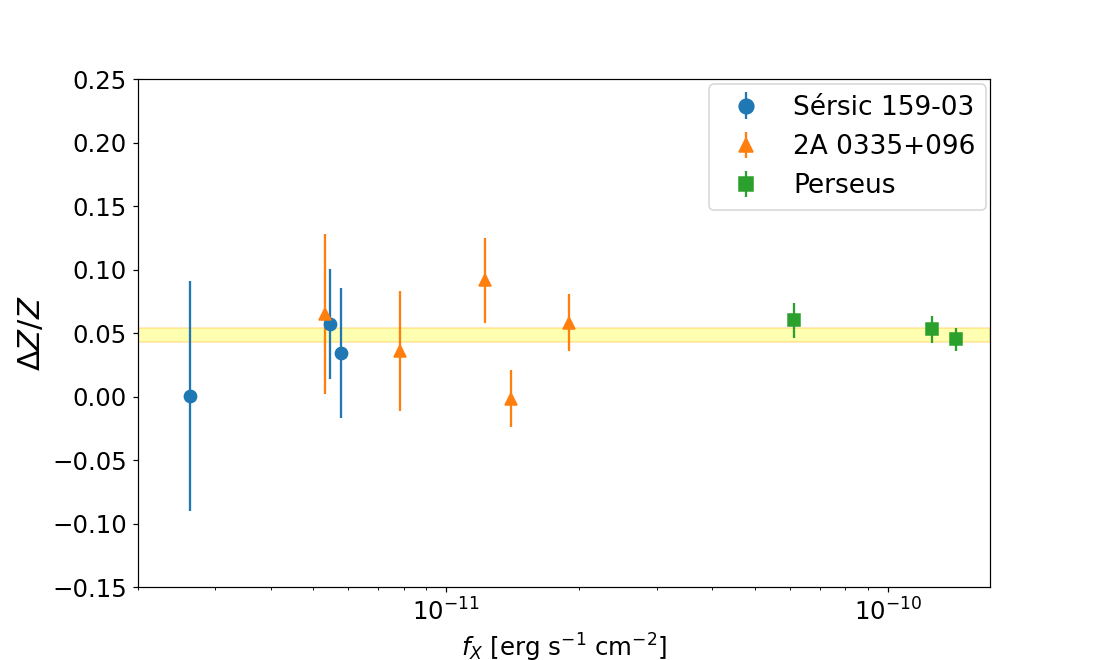}}
            \hfill
            \subfloat{\includegraphics[height=0.3\textwidth]{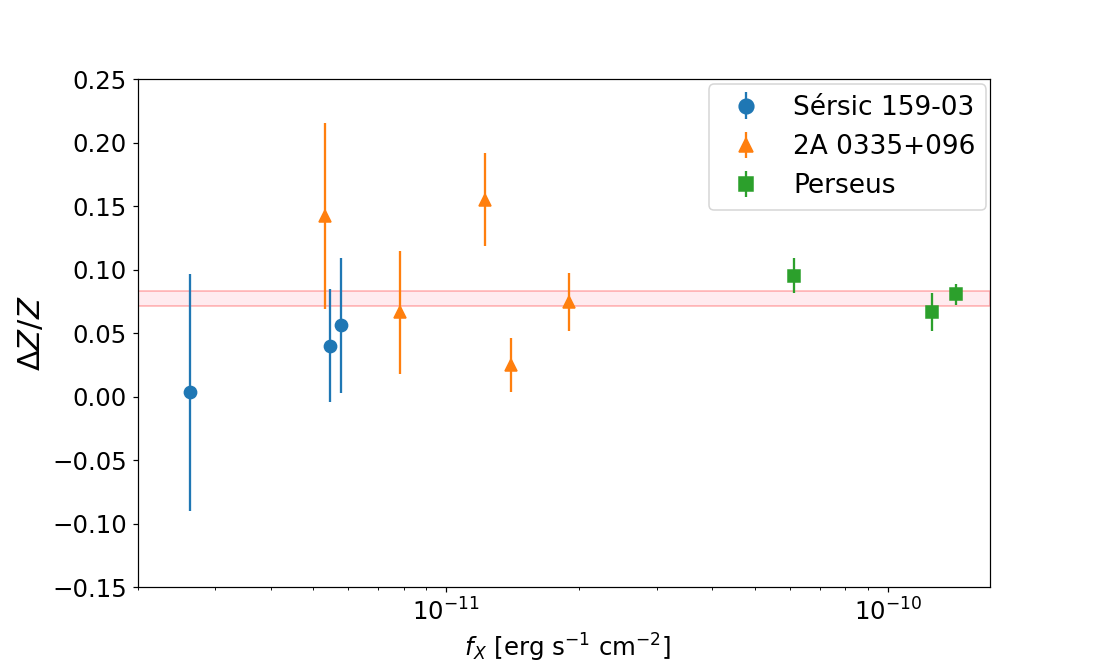}} \hfill
            \subfloat{\includegraphics[height=0.3\textwidth]{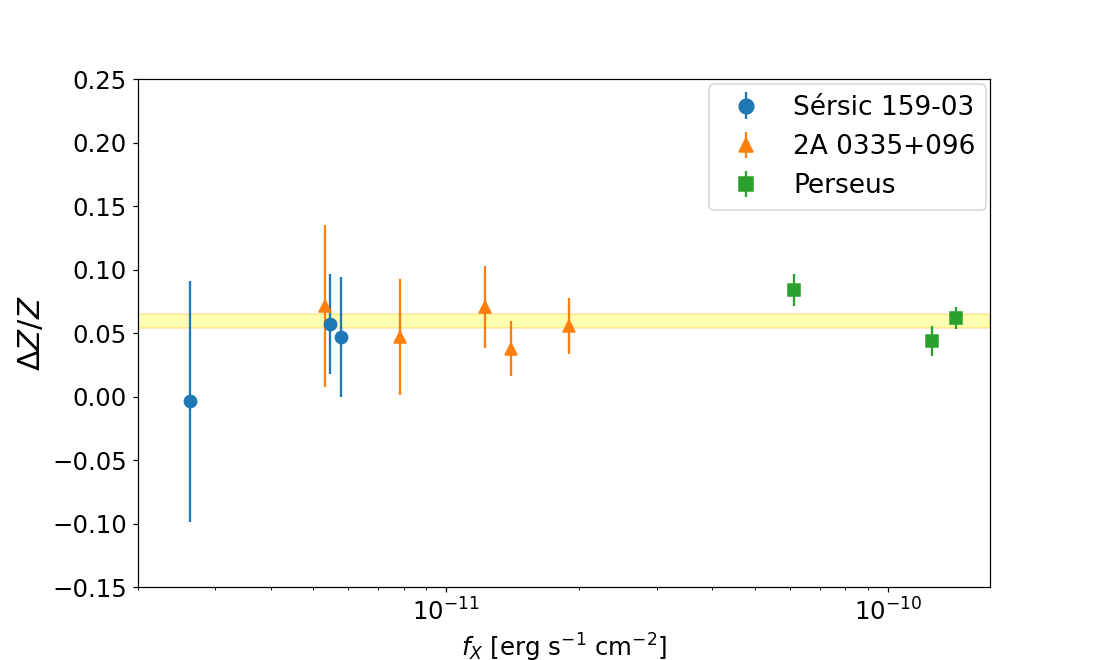}}
      \caption{Systematic error $\Delta Z/Z$ linked to the L-shell as a function of the X-ray flux. The division in the four panels is the same as in Fig. \ref{Temp}. The pink and the yellow regions represent the weighted averages of the results, together with their $1\sigma$ errors.}
         \label{Flux}
\end{figure*}
We also study a possible dependence of the $\Delta Z/Z$ values on the X-ray flux. In fact, in condition of high source count rates, additional systematic errors may be included in the analysis (systematics linked to the modelling, calibration issues of the instruments etc; see \citeauthor{mernier18a} \citeyear{mernier18a} for a review of the systematic errors that may affect the ICM abundances) and lead to an overestimate of the systematics related to the L-shell. 
The X-ray flux of each ring is calculated through the XSPEC \textit{flux} command\footnote{https://heasarc.gsfc.nasa.gov/xanadu/xspec/manual/node96.html.}  in the $0.7-10$ keV energy range (the same used in the spectral fitting) for the pn-camera spectra. The dependence of the systematics on the X-ray flux is shown in Fig. \ref{Flux}. As already discussed for Fig. \ref{Temp}, left and right panels of Fig. \ref{Flux} are obtained through the $1^{\text{st}}$ and the $2^{\text{nd}}$ strategies, respectively, while the top and bottom division refers to the different fitting option of the spectra (individual and joint fit of the MOS and pn spectra of each region, respectively). In Fig, \ref{Flux}, we also show the average $\Delta Z /Z$ values, already discussed in Sect. 4.1, as the pink and the yellow regions.

An inspection of Fig. \ref{Flux} shows no evident variations with the flux and the systematic error is consistent with being constant, within the statistical errors, over the considered flux range. In the low-flux range, large statistical errors do not allow to derive precise conclusions, while at higher flux statistical errors are tiny and systematics becomes dominant.

\section{Discussion}
In this work, we investigate systematic errors related to the Fe L-shell blend ($0.9-1.3$ keV energy range), which may affect iron-abundance measurements in intermediate-mass ($M_{500}=2-4\times10^{14}$ M$_{\odot}$, with temperature $T=2.5-4.5$ keV) galaxy clusters. The resolution of the currently available instrumentation, as well as our knowledge of the atomic processes, makes the modelling of the Fe L-shell particularly challenging, resulting in potential biases for our iron-abundance measurements. As already mentioned in the introduction, at temperatures higher than $\sim4$ keV, iron-abundance measurements are mostly driven by the K$\alpha$ emission at $\sim 6.7$ keV while, going down to the intermediate $2.5-4.5$ keV temperature range, the EW of the K$\alpha$ line reduces and the L-shell starts to assume a leading role in the fitting. 

\begin{figure}
   \includegraphics[height=0.3\textwidth]{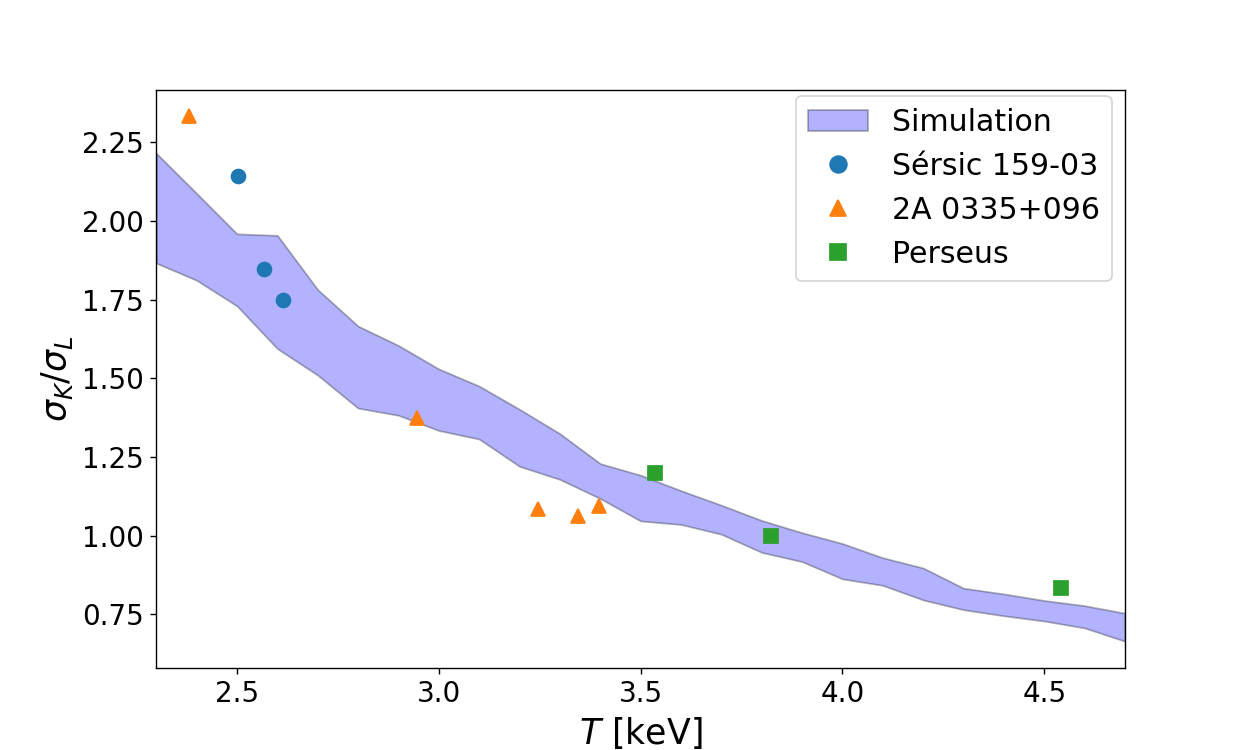}
   \caption{Ratio between statistical errors associated to the K$\alpha$- and L-shell-derived measurements, as a function of the ring temperatures, for the three clusters of our sample. Violet area represents $\sigma_K/\sigma_L$ values (with their $1\sigma$ scatter) derived through high-S/B simulated spectra, as explained in the text. Errors are estimated following the $2^{\text{nd}}$ strategy and through the joint fit of MOS and pn camera spectra of each region. For temperatures lower than $\sim3.5$ keV, the above ratio is larger than $1$, meaning that the L-shell drives the abundance measurements.}
   \label{errors}
\end{figure}

The increasing importance of the L-shell blend in the fitting is highlighted in Fig. \ref{errors}, where we show the ratio between the statistical errors associated to $Z_{\text{K}\alpha}$ and $Z_{\text{L-shell}}$ (Sect. 3.2.3) for the clusters in our sample, obtained through the $2^{\text{nd}}$ strategy (Sect. 3.2.3) and with a joint fit of the MOS and pn spectra of each region. 
To further highlight the interplay between the K$\alpha$ and the L-shell emissions at different temperatures, we also produced high-S/B simulated spectra, characterised by single temperature, fixed metallicity ($Z = 0.4$ Z$_{\odot}$) and normalisation equal to $0.1$ (the same for Perseus, i.e. one of the brightest clusters of the X-ray sky). We sampled the $2.3-4.7$ keV range every 0.1 keV, and for each step we produced 100 mock spectra with that temperature. By doing this, we were able to measure the median value of the simulated $\sigma_K/\sigma_L$ ratio (together with its $1\sigma$ uncertainty; violet region in Fig. \ref{errors}) for each temperature, through the $2^{\text{nd}}$ strategy and with a joint fit of MOS and pn spectra of each region.   
An inspection of Fig. \ref{errors} shows that our simulation manages to reproduce the measured $\sigma_K/\sigma_L$ trend. We found that, at temperatures lower than $\sim3.5$ keV, the $\sigma_K/\sigma_L$ ratio becomes larger than $1$, meaning that the L-shell blend has larger statistics than the K$\alpha$ and represents the leading line emission for the iron-abundance measurement.

The three clusters in our sample (Table \ref{table:1}) allow us to homogeneously study the intermediate $2.5-4.5$ keV range (Fig. \ref{Temp}), particularly in condition of negligible background emission, and so to explore a temperature range where the transition between K$\alpha$- and L-shell-driven fits is observed. In such a favourable situation, where the K$\alpha$ and the L-shell emissions have similar EWs, to estimate the systematics linked to the modelling of the L-shell, we can simply compare the L- to the K-derived iron-abundance measurements, since the latter are currently considered reliable, as previously discussed. In the case of an individual fit of each detector spectrum, the L-shell is found to lead to $\sim 5-6\%$ systematics (Fig. \ref{Temp}, top), in the whole $2.5-4.5$ keV temperature range. We point out that this is actually an upper limit to the systematic error linked to the modelling of the L-shell, since this value may include the contribution of other systematics. Conversely, the joint fits of the three detectors lead to $\sim6-8\%$ systematics (Fig. \ref{Temp}, bottom), a $\sim1-2\%$ excess which is likely due to the cross-calibration issues of the EPIC detectors, as already discussed in Sects. 3.2.3 and 4.1. The strategies we have adopted in Sect. 3.2.3 represent the most-straightforward way to quantify the L-blend-induced systematics and, thanks to a carefully selected sample of galaxy clusters, they lead us to the most accurate (and precise) results that can be obtained with the currently available instrumentation. Indeed, in our work we carefully focus on a sample of clusters with long XMM-\textit{Newton} observations and high flux in the $0.1-2.4$ keV band (Table \ref{table:1}), at low redshifts ($z\sim 0.018-0.056$) and in the $2.5-4.5$ keV temperature range. As already discussed, such conditions allowed us to easily compare K$\alpha$- and L-shell-derived measurements.

When performing abundance measurements on galaxy clusters, both the K$\alpha$ and L-shell emission bands are generally used in the fitting, to maximise the available statistic. Thus the measured $5-8\%$ value must be considered as an upper limit to the contribution of the L-shell-induced systematics to the derived iron abundances. This consideration gives us confidence for future measurements in intermediate-mass ($M_{500}= 2-4 \times 10^{14}$ M$_{\odot}$) galaxy clusters, at least in low-background-emission regions. In addition, when studying individual galaxy clusters, the precision of the results is usually limited by the poor data quality, with the statistical errors associated to the measured abundances being larger than the derived $5-8\%$ systematics. This means that abundance measurements are usually mostly influenced by the low statistics, while systematics do not significantly affect the results, making the abundance measurements reliable. However, systematic errors will be dominant when averaging abundance measurements over a sufficiently large sample of clusters. For this reason, when performing iron-abundance measurements over a large sample of intermediate-mass clusters with temperatures lower than $\sim3.5$ keV, where the L-shell becomes the leading emission, the L-shell-induced systematics should no longer be neglected.

A detailed analysis of the L-shell systematics is of primary importance for future iron-abundance measurements in galaxy-cluster outskirts. As mentioned above, cluster external regions provide important clues on the physical properties of the ICM, and so they have recently enjoyed particular attention. \citet{ghizzardi21, ghizzardi21-corrigendum} focussed on the outskirts of massive clusters, but no reliable measurements have been carried out in the intermediate-mass (and temperature) range so far. These temperatures, together with the increasing background contribution which rises in the cluster outskirts, make the L-shell the leading emission for iron-abundance measurements. For this reason, an accurate characterisation of the systematics related to the L-shell is required before
proceeding with future iron-abundance measurements. In the cluster outskirts, additional systematics (particularly related to the background emission) may influence the measured abundances. However, our work places a limit on the L-shell-induced systematics and represents a good starting point for future iron-abundance measurements.

Future X-ray missions, such as XRISM (\citeauthor{tashiro18} \citeyear{tashiro18}; \citeauthor{guainazzi18} \citeyear{guainazzi18}) and ATHENA (\citeauthor{barret13} \citeyear{barret13}; \citeauthor{cucchetti19} \citeyear{cucchetti19}), will carry microcalorimeters and this will allow to perform high-resolution spectroscopy, giving the opportunity to test, with unprecedented detail, the atomic codes underlying iron-abundance measurements. Systematic errors related to the spectral resolution of the instrumentation will be less important, while those related to the theoretical models which describe the emission lines will become dominant.
\section{Summary}
We estimated a systematic error related to the modelling of the Fe L-shell that affects iron-abundance measurements in the $2.5-4.5$ keV temperature range. The main results of our work are the following:
\begin{itemize}
    \item the analysis carried out on three clusters shows that L-induced systematics range between $5-8\%$, depending on the strategy and on the spectral fitting option (Fig. \ref{2A}). This result indicates that systematics have a modest contribution to the measured abundances;
    \item considering MOS and pn spectra of each region individually in the fitting leads to $5-6\%$ systematics, in the whole $2.5-4.5$ keV temperature range (Fig. \ref{Temp}, top); 
    \item a joint fit of the MOS and pn spectra of each region leads on average to a $1-2\%$ excess in the estimate of the systematics, with respect to the individual fits (Fig. \ref{Temp}, bottom);
    \item the systematics show no evident dependence on the X-ray flux (Fig. \ref{Flux}).
\end{itemize}
Finally and especially when averaging iron abundances over a sufficiently large sample of clusters with temperatures lower than $\sim 3.5$ keV, we suggest to take into account the measured $5-8\%$ systematic contribution related to the L-shell to the derived abundances, in order to correct L-shell-induced systematics.

\begin{acknowledgements}
      We thank the anonymous referee for useful suggestions. GR acknowledges funding from Istituto Nazionale di Astrofisica (INAF) and from the AHEAD project (grant agreement n. 654215), which is part of the EU-H2020 programme. We acknowledge financial support contributions from contract n. 4000116655/16/NL/BW by ESA and from contract  ASI-INAF n. 2019-27-HH.0.
      The results reported in this article are based on data obtained from the XMM-\textit{Newton} observatory, an ESA science mission with instruments and contributions directly funded by ESA Member States and USA (NASA).
\end{acknowledgements}

%
%


\bibliographystyle{aa} 
\bibliography{biblio.bib} 


\begin{appendix}
\section{Iron-abundance measurements}

In Table \ref{individual} and Table \ref{joint}, we report the derived iron abundances for Perseus, 2A 0335+096 and Sérsic 159-03, measured after the exclusion of the L-shell ($Z_{\text{K}\alpha}$) and the K$\alpha$ ($Z_{\text{L-shell}}$) and obtained either through the individual (Table \ref{individual}) and the joint (Table \ref{joint}) fit of MOS and pn spectra.

\renewcommand\arraystretch{1.25}
\renewcommand{\tabcolsep}{5.5pt}
\begin{strip}
    \captionof{table}{\label{individual} Iron-abundance values measured for Perseus, 2A 0335+096 and Sérsic 159-03 after the exclusion of the L-shell ($Z_{\text{K}\alpha}$) and the K$\alpha$ ($Z_{\text{L-shell}}$) and by individually fitting MOS and pn spectra of each region.}
	    
		\centering
		\small
		\begin{tabular}{ c|c|c|ccc|ccc }
        \hline
        \hline
         Cluster&Strategy &Rings ( $'$) &\multicolumn{3}{c}{\textit{Z}$_{\text{K}\alpha}$  (Z$_{\odot}$)} & \multicolumn{3}{c}{\textit{Z}$_{\text{L-shell}}$  (Z$_{\odot}$)} \\
         \cline{4-9}
        &&&MOS 1 & MOS 2 & pn &MOS 1 & MOS 2 & pn\\
        \hline
        &&$0.5-1$& $0.695\pm0.012$ & $0.636\pm0.011$ & $0.582\pm0.009$ & $0.714\pm0.010$ & $0.683\pm0.021$ & $0.683\pm0.008$\\
        &$1$ &$1-2$& $0.723\pm0.009$ & $0.659\pm0.008$ & $0.650\pm0.006$ & $0.758\pm0.008$ & $0.710\pm0.014$ & $0.691\pm0.006$\\
        Perseus &&$2-3$& $0.697\pm0.010$ & $0.662\pm0.009$ & $0.630\pm0.007$ & $0.763\pm0.011$ & $0.652\pm0.016$ & $0.654\pm0.007$\\
        \cline{2-9}
        &&$0.5-1$& $0.698\pm0.012$ & $0.655\pm0.012$ & $0.590\pm0.010$ & $0.713\pm0.010$ & $0.695\pm0.010$ & $0.654\pm0.011$\\
        &$2$ &$1-2$& $0.744\pm0.010$ & $0.680\pm0.008$ & $0.661\pm0.007$ & $0.750\pm0.007$ & $0.718\pm0.008$ & $0.704\pm0.006$\\
        &&$2-3$& $0.719\pm0.010$ & $0.664\pm0.009$ & $0.629\pm0.007$ & $0.757\pm0.008$ & $0.679\pm0.012$ & $0.672\pm0.007$\\
        \hline
        &&$0.5-1$& $0.706\pm0.027$ & $0.701\pm0.027$ & $0.726\pm0.019$ & $0.676\pm0.016$ & $0.680\pm0.018$ & $0.763\pm0.030$\\
        &&$1-2$& $0.628\pm0.023$ & $0.649\pm0.022$ & $0.632\pm0.015$ & $0.668\pm0.027$ & $0.699\pm0.029$ & $0.638\pm0.027$\\
        &$1$ &$2-3$& $0.573\pm0.058$ & $0.536\pm0.026$ & $0.566\pm0.019$ & $0.675\pm0.038$ & $0.612\pm0.036$ & $0.623\pm0.018$\\
        &&$3-4$& $0.453\pm0.038$ & $0.572\pm0.035$ & $0.520\pm0.025$ & $0.520\pm0.029$ & $0.575\pm0.045$ & $0.549\pm0.025$\\
        2A 0335+096&&$4-5$& $0.573\pm0.049$ & $0.498\pm0.044$ & $0.443\pm0.029$ & $0.547\pm0.049$ & $0.509\pm0.055$ & $0.546\pm0.032$\\
        \cline{2-9}
        &&$0.5-1$& $0.706\pm0.027$ & $0.696\pm0.026$ & $0.728\pm0.019$ & $0.677\pm0.022$ & $0.680\pm0.017$ & $0.745\pm0.011$\\
        &&$1-2$& $0.628\pm0.023$ & $0.645\pm0.022$ & $0.632\pm0.015$ & $0.660\pm0.021$ & $0.689\pm0.021$ & $0.668\pm0.011$\\
        &$2$ &$2-3$& $0.544\pm0.028$ & $0.536\pm0.026$ & $0.568\pm0.021$ & $0.618\pm0.026$ & $0.593\pm0.023$ &  $0.604\pm0.017$\\
        &&$3-4$& $0.453\pm0.034$ & $0.572\pm0.035$ & $0.525\pm0.025$ & $0.507\pm0.032$ & $0.569\pm0.034$ &  $0.539\pm0.025$\\
        &&$4-5$& $0.569\pm0.050$ & $0.498\pm0.045$ & $0.448\pm0.032$ & $0.543\pm0.042$ & $0.507\pm0.038$ &  $0.527\pm0.029$\\
        \hline
        &&$0.5-1$& $0.452\pm0.038$ & $0.433\pm0.031$ & $0.403\pm0.025$ & $0.433\pm0.024$ & $0.480\pm0.023$ & $0.415\pm0.012$\\
        &$1$ &$1-2$& $0.331\pm0.030$ & $0.337\pm0.032$ & $0.365\pm0.021$ & $0.332\pm0.025$ & $0.365\pm0.019$ & $0.343\pm0.014$\\
        Sérsic 159-03&&$2-3$& $0.273\pm0.043$ & $0.266\pm0.040$ & $0.308\pm0.035$ & $0.249\pm0.023$ & $0.333\pm0.043$ & $0.291\pm0.022$\\
        \cline{2-9}
        &&$0.5-1$& $0.455\pm0.033$ & $0.429\pm0.031$ & $0.398\pm0.021$ & $0.433\pm0.021$ & $0.485\pm0.024$ & $0.429\pm0.011$\\
        &$2$ &$1-2$& $0.332\pm0.029$ & $0.332\pm0.028$ & $0.352\pm0.022$ & $0.334\pm0.021$ & $0.367\pm0.020$ & $0.357\pm0.011$\\
        &&$2-3$& $0.271\pm0.042$ & $0.267\pm0.042$ & $0.315\pm0.036$ & $0.255\pm0.023$ & $0.317\pm0.025$ & $0.297\pm0.018$\\
        \hline
        \end{tabular}
	
\end{strip}


\begin{strip}
     
	    \captionof{table}{Same as Table \ref{individual}, with the difference that MOS and pn spectra are jointly fitted.}
		\centering
		\begin{tabular}{ c|c|cc|cc }
        \hline
        \hline
          Cluster& Rings ( $'$)&\multicolumn{2}{c}{$1^{\text{st}}$ strategy} & \multicolumn{2}{c}{$2^{\text{nd}}$ strategy} \\
        \cline{3-6}
        &&\textit{Z}$_{\text{K}\alpha}$ (Z$_{\odot}$) & \textit{Z}$_{\text{L-shell}}$ (Z$_{\odot}$) & \textit{Z}$_{\text{K}\alpha}$ (Z$_{\odot}$) & \textit{Z}$_{\text{L-shell}}$ (Z$_{\odot}$) \\
        \hline
        &$0.5-1$ & $0.618\pm0.006$ & $0.677\pm0.006$ & $0.620\pm0.006$ & $0.672\pm0.005$\\
        Perseus&$1-2$ & $0.669\pm0.004$ & $0.723\pm0.004$ & $0.677\pm0.004$ & $0.719\pm0.004$\\
        &$2-3$ & $0.643\pm0.005$ & $0.686\pm0.008$ & $0.663\pm0.005$ & $0.692\pm0.006$\\
        \hline
        &$0.5-1$ & $0.713\pm0.013$ & $0.731\pm0.008$ & $0.711\pm0.014$ & $0.738\pm0.006$\\
        &$1-2$ & $0.628\pm0.012$ & $0.675\pm0.008$ & $0.628\pm0.011$ & $0.663\pm0.008$\\
        2A 0335+096&$2-3$ & $0.548\pm0.015$ & $0.633\pm0.013$ & $0.552\pm0.013$ & $0.591\pm0.012$\\
        &$3-4$ & $0.511\pm0.018$ & $0.545\pm0.017$ & $0.512\pm0.017$ & $0.536\pm0.016$\\
        &$4-5$ & $0.485\pm0.024$ & $0.554\pm0.026$ & $0.488\pm0.023$ & $0.523\pm0.021$\\
        \hline
        &$0.5-1$ & $0.423\pm0.016$ & $0.440\pm0.010$ & $0.421\pm0.015$ & $0.445\pm0.007$\\
        Sérsic 159-03&$1-2$ & $0.338\pm0.016$ & $0.357\pm0.008$ & $0.341\pm0.014$ & $0.357\pm0.008$\\
        &$2-3$ & $0.285\pm0.022$ & $0.286\pm0.015$ & $0.288\pm0.024$ & $0.287\pm0.013$\\
        \hline
        \end{tabular}
	\label{joint}
\end{strip}

\cleardoublepage
\section{Perseus and Sérsic 159-03}

\begin{figure}[h!]
   \subfloat{\includegraphics[height=0.3\textwidth]{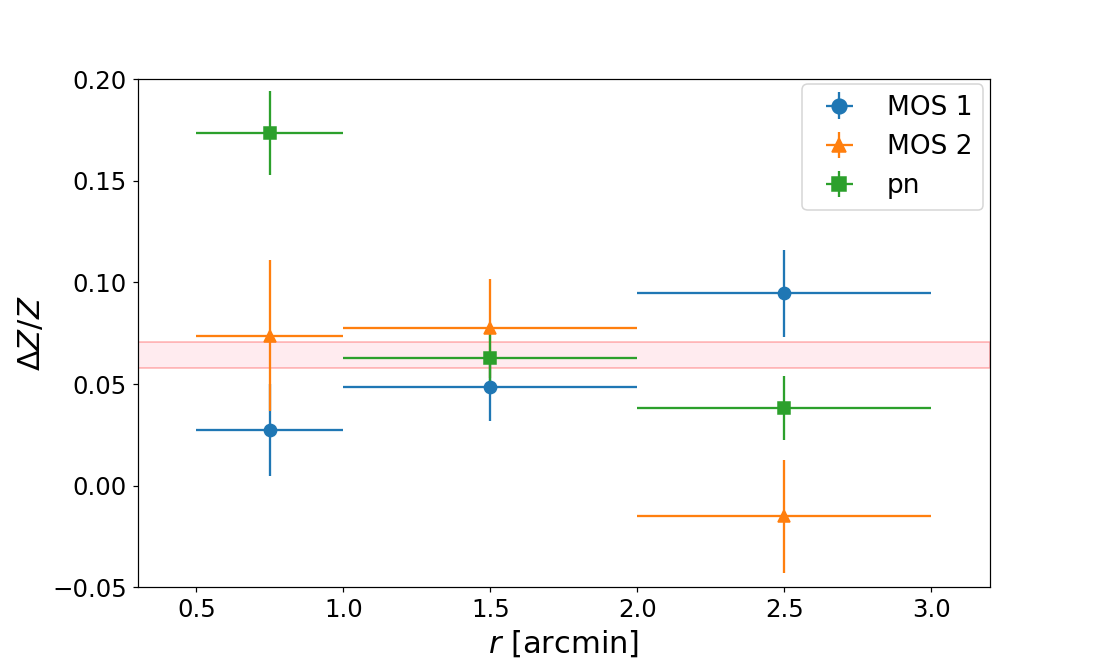}} \hfill
   \subfloat{\includegraphics[height=0.3\textwidth]{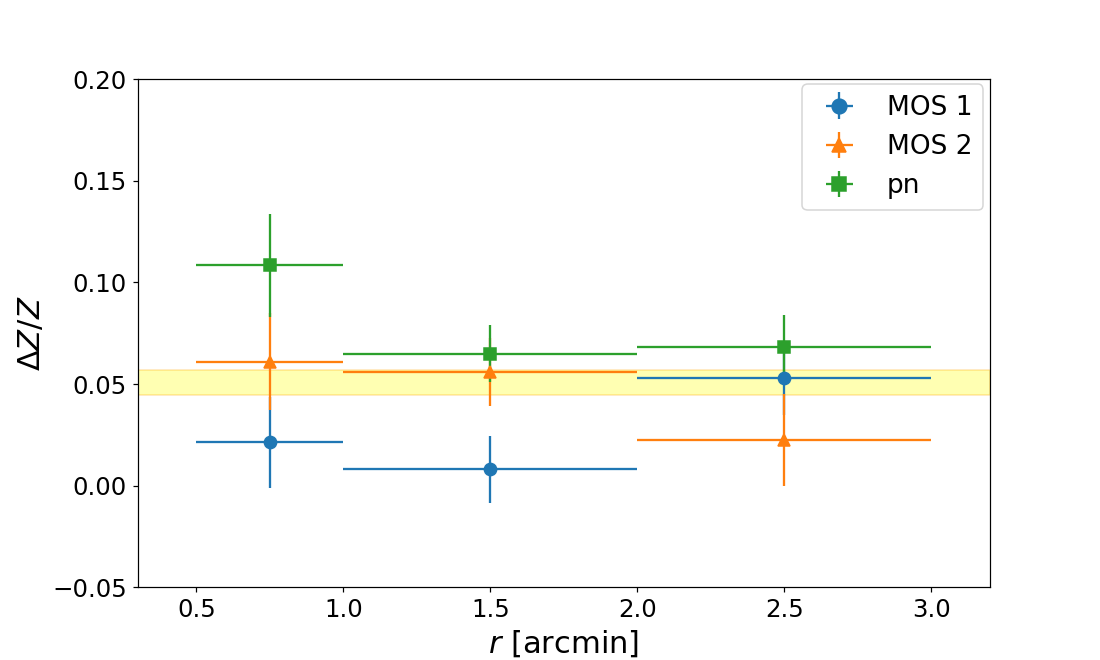}}\hfill
   \subfloat{\includegraphics[height=0.3\textwidth]{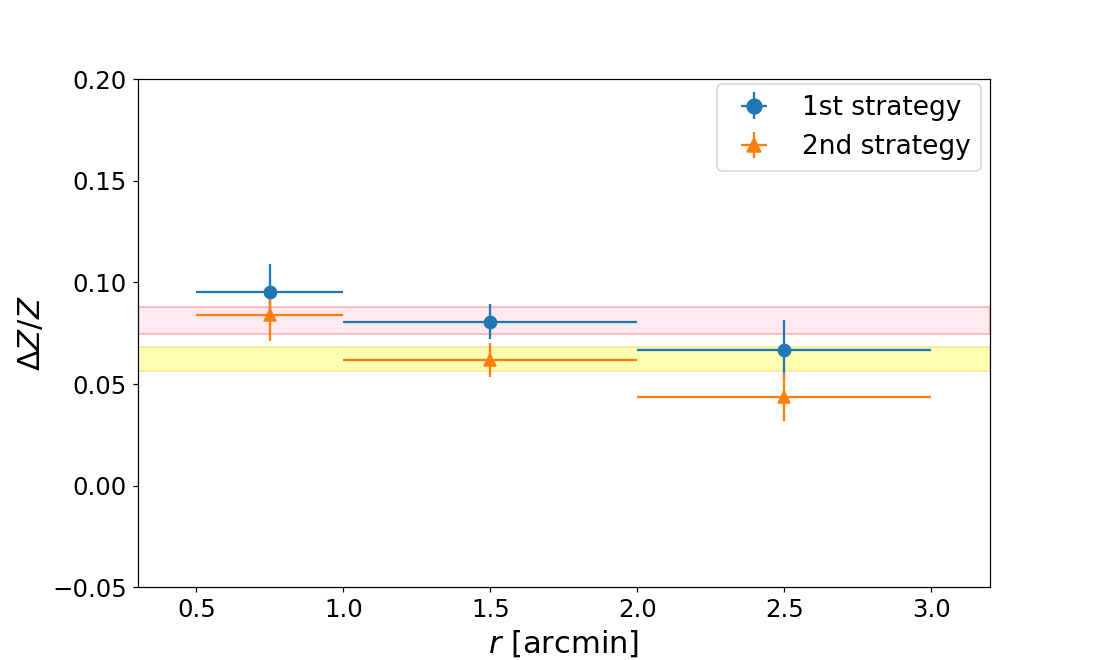}}
   \caption{Radial profile of the systematic error $\Delta Z/Z$ for Perseus. In the top and the centre panels, we show the results obtained through an individual fit of the MOS and pn spectra of each region, following the $1^{\text{st}}$ and the $2^ {\text{nd}}$ strategies, respectively. In the bottom panel, we show the same profiles, with the difference that MOS and pn spectra are considered jointly in the fitting. The weighted averages of the results are shown, together with their $1\sigma$ errors, as the pink and the yellow regions, for the $1^{\text{st}}$ and the $2^ {\text{nd}}$ strategies, respectively.}
   \label{P}
\end{figure}

\begin{figure}[h!]
   \subfloat{\includegraphics[height=0.3\textwidth]{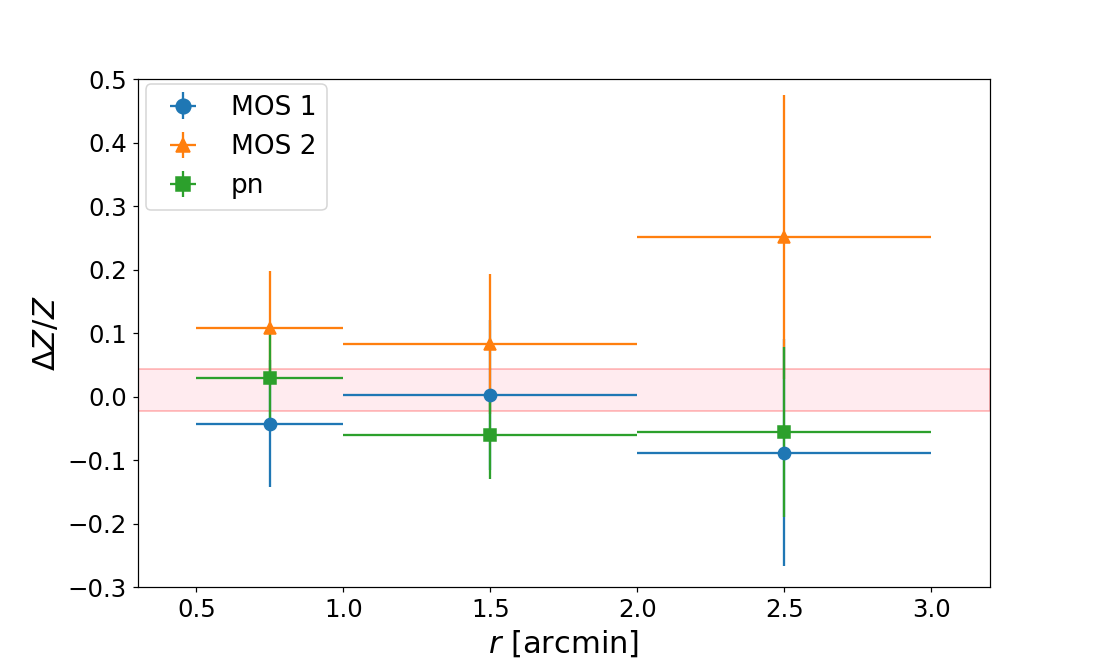}} \hfill
   \subfloat{\includegraphics[height=0.3\textwidth]{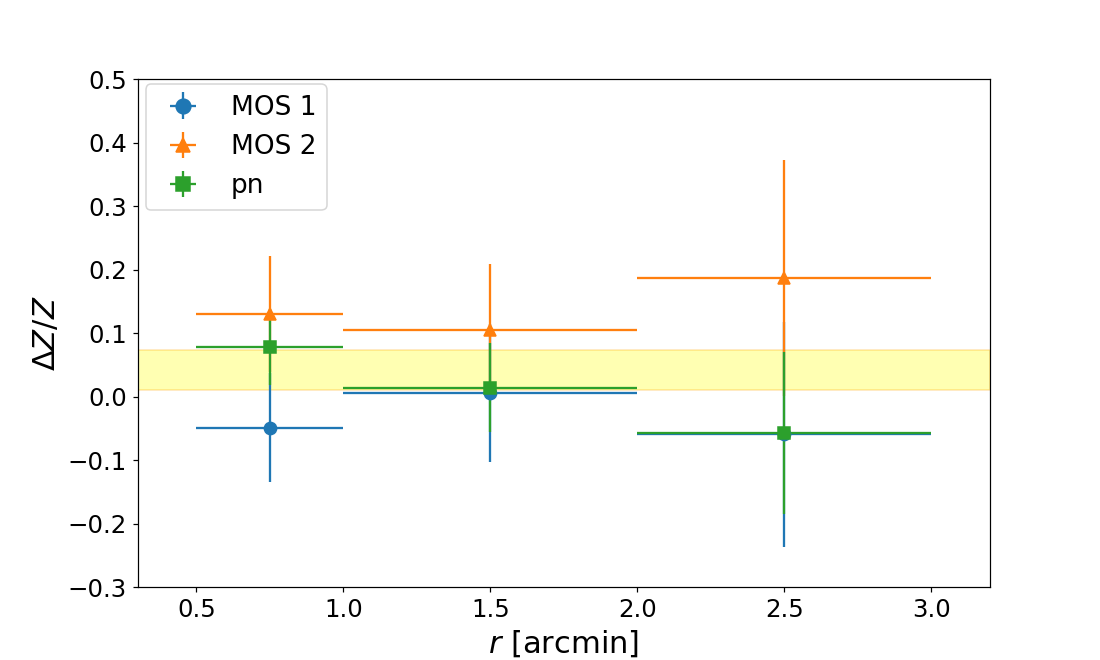}}\hfill
   \subfloat{\includegraphics[height=0.3\textwidth]{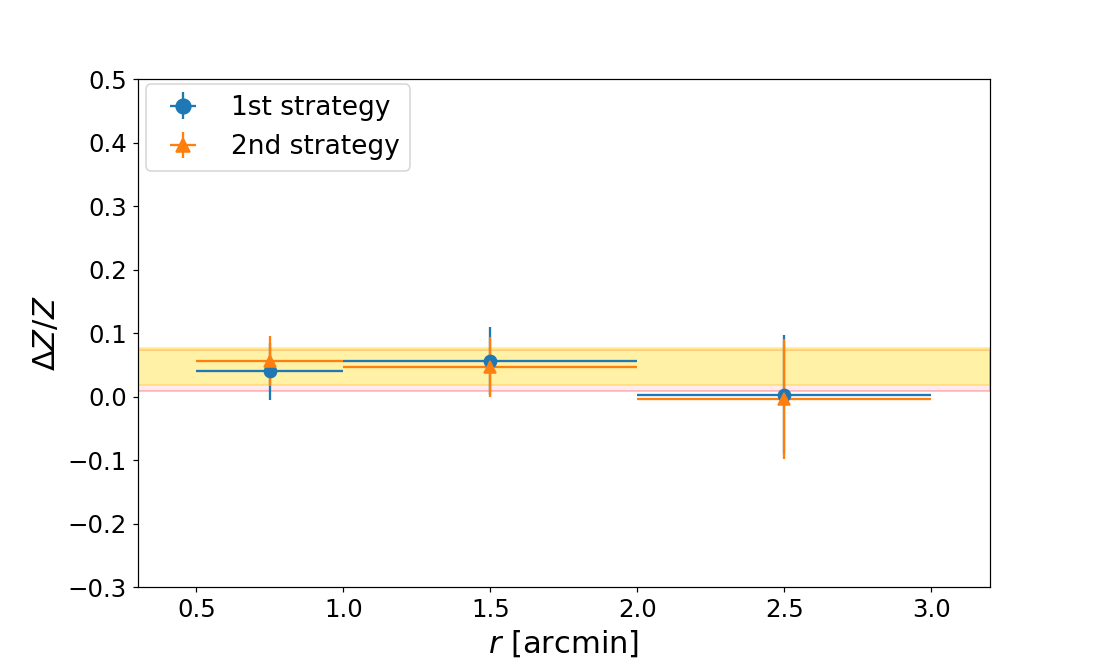}}
   \caption{Same as Fig. \ref{P}, for Sérsic 159-03.}
   \label{S}
\end{figure}

In Fig. \ref{P} and Fig. \ref{S}, we show the derived $\Delta Z/Z$, as a function of the cluster radius, for Perseus and Sérsic 159-03, respectively. $\Delta Z/Z$ values are obtained either by individually fitting each detector spectrum and from a joint fit.

\section{Temperature values}
In Table \ref{tab_temp}, we report the measured temperatures for Perseus, 2A 0335+096 and Sérsic 159-03, by individually fitting MOS and pn spectra of each region, which is a useful sanity check. For spectra that were better described by the double-temperature model (see Sect. 3.2.2), we report a weighted-averaged temperature on the normalisations of the APEC+APEC model, calculated as follows:
\begin{equation}
    T = \frac{T_{\text{cold}}\times \text{norm}_{\text{cold}} + T_{\text{hot}}\times \text{norm}_{\text{hot}}}{\text{norm}_{\text{cold}} + \text{norm}_{\text{hot}}},
\end{equation}
where $T_\text{cold}$ and $T_\text{hot}$ are the lowest and the highest derived temperatures through the 2T 1Z model, while $\text{norm}_{\text{cold}}$ and $\text{norm}_{\text{hot}}$ are their normalisations, respectively.

 \renewcommand\arraystretch{1.25}
 \renewcommand{\tabcolsep}{3pt}
\begin{table}
    \captionof{table}{\label{tab_temp} Temperature values measured for Perseus, 2A 0335+096 and Sérsic 159-03 by individually fitting MOS and pn spectra of each region.}
	    
		\centering
		\small
		\begin{tabular}{ c|c|ccc}
        \hline
        \hline
        Cluster&Rings ( $'$) &\multicolumn{3}{c}{Temperature  (keV)}\\
        \cline{3-5}
        &&MOS 1 & MOS 2 & pn \\
        \hline
        &$0.5-1$& $3.924\pm0.087$ & $3.933\pm 0.119$ & $3.696 \pm 0.055$ \\
        Perseus&$1-2$& $4.369\pm0.088$ & $4.161\pm0.073$ & $4.042\pm0.036$ \\
        &$2-3$& $5.212\pm0.165$ & $4.900\pm0.226$ & $4.671\pm0.072$ \\
        \hline
        &$0.5-1$& $2.565 \pm 0.124$ & $2.539\pm 0.156$&  $2.431 \pm 0.047$\\
        &$1-2$& $3.248\pm0.288$ & $3.266\pm0.428$& $3.039\pm0.138$ \\
        2A 0335+096& $2-3$&  $3.326\pm0.024$& $3.230\pm0.033$ & $3.198\pm0.020$  \\
        &$3-4$& $3.430\pm0.055$ & $3.439 \pm 0.042$& $3.347 \pm 0.023$ \\
        &$4-5$& $3.493\pm0.078$ & $3.390 \pm 0.068$ & $3.357\pm0.052$ \\
        \hline
        &$0.5-1$& $2.750\pm0.353$ & $2.532\pm0.103$ & $2.688\pm0.187$ \\
        Sérsic 159-03&$1-2$& $2.768\pm0.414$ & $2.666\pm0.213$& $2.729\pm0.345$ \\
        &$2-3$& $2.564\pm0.034$ & $2.536\pm0.039$& $2.585\pm0.028$ \\
        \hline
        \end{tabular}
        \tablefoot{For cluster rings that were better described by the 2T 1Z model, we report a weighted-averaged temperature on the normalisations of the APEC+APEC model, as described in the text.}
	
\end{table}

\end{appendix}

\end{document}